\documentclass{jfm}
\usepackage{epsfig,psfrag}
\usepackage{amsmath}
\usepackage{amsfonts}
\usepackage{mathrsfs}
\usepackage{amssymb}
\usepackage{bm}
\usepackage[usenames,dvipsnames]{color}
\usepackage{natbib}
\usepackage{mdwlist}
\usepackage{xr,url}

\begin{document}

\newcommand\beq{\begin{equation}}
\newcommand\eeq{\end{equation}}
\newcommand\beqa{\begin{eqnarray}}
\newcommand\eeqa{\end{eqnarray}}
\newcommand{\Sy}{{\cal S}}
\newcommand{\U}{{\cal U}}
\newcommand{\K}{{\cal K}}
\newcommand{\T}{\mathsfi {T}} % Schmidt number
\newcommand{\bx}{\mathbf{x}}
\newcommand{\by}{\mathbf{y}}
\newcommand{\bz}{\mathbf{z}}
\newcommand{\hx}{\mathbf{ \hat{x}}}
\newcommand{\hy}{\mathbf{ \hat{y}}}
\newcommand{\hz}{\mathbf{ \hat{z}}}
\newcommand{\hg}{\mathbf{ \hat{g}}}
%%%%%%%%%%%%%%%%%%%%%%%%%%%%%%%%%%%%%%%%%%%%%%%%%%%%%%%%%%%%%%%%%%%%%%%%

\def\half{\frac{1}{2}}
\def\quart{\frac{1}{4}}
\def\ud{{\rm d}}
\def\eps{\epsilon}

\title[Layer formation in horizontally forced stratified turbulence]{Layer formation in horizontally forced stratified turbulence: connecting exact coherent structures to linear instabilities}

\author{Dan Lucas\aff{1}\corresp{\email{d.lucas1@keele.ac.uk}} C. P. Caulfield\aff{2,3} \& Rich R.  Kerswell\aff{4} }
\affiliation{\aff{1}{School of Computing and Mathematics, Keele University, Staffordshire, ST5 5BG}\aff{2}{Department of Applied Mathematics \& Theoretical Physics, University of Cambridge, Centre for Mathematical Sciences, Wilberforce Road, Cambridge, CB3 0WA, UK}
\aff{3}{BP Institute, University of Cambridge, Madingley Rise, Madingley Road, Cambridge, CB3 0EZ, UK}\aff{4}{School of Mathematics, University of Bristol, University Walk, Bristol, BS8 1TW, UK.}}

\maketitle
\begin{abstract}
We consider turbulence in a stratified `Kolmogorov' flow, driven by  horizontal shear in the form of sinusoidal body forcing in the presence of an imposed  background linear stable stratification in the third direction. This flow configuration allows the controlled investigation of the formation of coherent structures, which here organise the flow into horizontal layers by inclining the background shear as the strength of the stratification is increased.  By numerically converging exact steady states from direct numerical simulations of chaotic flow, we show, for the first time, a robust connection between  linear theory predicting instabilities from infinitesimal perturbations  to the robust finite amplitude nonlinear layered state observed in the turbulence. We investigate how the observed vertical length scales  are related to the primary linear instabilities and compare to  previously considered examples of shear instability leading to layer formation in other horizontally sheared flows.
% (Deloncle \emph{et al. J. Fluid Mech.}, 2007 vol. 570, pp. 297-305 Billant \& Chomaz \emph{J. Fluid Mech.} 2000 vol. 419, pp. 65-91.)

 \end{abstract}
%%%%%%%%%%%%%%%%%%%%%%%%%%%%%%%%%%%%%%%%%%%%%%%%%%%%%%%%%%%%%%%%%%%%%%%%

\section{Introduction}

Laboratory experiments, numerical simulations and even field measurements of turbulent flows which are strongly stably stratified are regularly observed to exhibit  spontaneous layering where the density field is organised into relatively deep, relatively well-mixed regions or `layers' separated by relatively thin `interfaces' with enhanced density gradients \citep{Park1994, Holford:1999uc, 1999DyAtO..30..173H,Oglethorpe:2013cv, Thorpe:2016ga,2016JPO....46.1023F,Leclercq:2214839}. Scaling analyses  \citep{Billant:2001cs,LINDBORG:2006eg} have provided some theoretical basis for the expected behaviour of vertical `layer' scales relative to the basic parameters involved in the distinguished asymptotic limit of extremely `strong' stratification and intense turbulence. At its heart this scaling has the central idea that sufficiently strong stratification 
(definable in a precise fashion) inevitably introduces anisotropy into the velocity field: vertical velocities are suppressed by the buoyancy force compared to horizontal velocity components, thus leading to pancake-like layering, with characteristic turbulent regions
having much larger horizontal extent $l_h$ than vertical extent $l_v$. Indeed, following
\cite{2016JPO....46.1023F}, we refer to this regime as the `layered anisotropic stratified turbulence' (LAST) regime.

Even in a purely one-dimensional model, where 
there is no characteristic horizontal scale $l_h$, and so the LAST regime is formally not
possible, the
hypothesis that sufficiently strong stratification  suppresses 
vertical motions can lead to a prediction of layering in the density field.
Specifically, if the vertical velocity is suppressed, it is at least 
plausible that some appropriately averaged 
vertical (turbulent) buoyancy flux should decrease
with sufficiently strong stratification. As originally argued by \cite{Phillips:1972ch}, if there is a range of stratifications for which the  vertical (turbulent) buoyancy flux decreases with increasing stratification, local perturbations in density gradient will tend to be intensified rather than smoothed out by turbulent mixing, suggesting that uniform density gradients are `unstable', in that they are prone to developing a layer-interface structure (see \cite{Park1994} for a clear discussion).
 Although in its simplest formulation (where for sufficiently strong stratification the buoyancy flux decreases monotonically with stratification) this Phillips mechanism is ill-posed, corresponding essentially to an  `anti-diffusive' problem, various regularisation mechanisms 
to  limit the `sharpness' of the interfacial density gradients have been proposed.  For example, \cite{Barenblatt1993} demonstrated that the underlying problem could become well-posed if there was a time-lag between the turbulence and the mixing irreversibly modifying the density distribution, and there
is at least some evidence that just such a time-lag exists
in transient turbulent mixing driven by shear instabilities \citep{Mashayek2013}. Alternatively,
\cite{Balmforth1998} proposed
that the relationship between buoyancy flux and stratification
should be `N-shaped', with a return to an increase in buoyancy flux with increasing and sufficiently large stratification. Indeed, the possibly non-monotonic dependence of irreversible buoyancy flux 
on external parameters is a very active area of  research controversy (see e.g \cite{Venayagamoorthy2016, Venaille2016, Maffioli2016}). Importantly however, the fundamental physical mechanisms leading to either the formation or the maintenance of layered density distributions are still quite open. 

One highly promising possible mechanism is suggested by the linear instability of a vortex dipole in a stratified environment, known as the zig-zag instability \citep{Billant:2000jg}.  The {\bf linear} theory of the instability provides a scaling for layer depth which has been confirmed {\bf at finite amplitude} numerically and experimentally \citep{Billant:2000df,Billant:2000jg,Billant:2000wg,2008JFM...599..229D,Waite:2008er,Augier:2015fo} resulting in the zig-zag instability being a popular explanation for the observation of layers \citep{Thorpe:2016ga}.
A significant question is therefore how generic is a `zig-zag' mechanism? Specifically, given other horizontally varying base flows, constituting 
other, less precisely organised distributions of vertical vorticity,  do analogous linear instabilities exist to provide vertical structure? In addition, is it possible to make a more robust connection between such a linear stability mechanism and a highly nonlinear, yet identifiable sustained turbulent state? For example \cite{2008JFM...599..229D} did not observe nonlinear saturation in their simulations of the zig-zag instability of counter-rotating vortex pairs. 

Here we consider the case when forcing provides a horizontal shear which resembles  in at least some respects the case of (vertically) stratified Taylor-Couette flow considered by \cite{Oglethorpe:2013cv,Leclercq:2214839} (minus rotation and curvature, but where both non-monotonic buoyancy flux with stratification and spontaneous layer formation is known to occur) and the vertically invariant base flows of \cite{Billant:2000jg} and \cite{DELONCLE:2007jl}. We are further motivated by the results
of \cite{Basak:2006iz} who considered the freely decaying case of a horizontal shear layer in a stratified environment. They found that, compared to its vertically sheared counterpart, the flow  exhibits more intense turbulence as the stratification does not penalise the initial two-dimensional linear instability of the shear layer. They also observed `dislocated pancake vortices', in that  the flow exhibits vertical structure, but low vertical velocity, at strong stratification, precisely 
as postulated for the Billant \& Chomaz scaling and the Phillips mechanism. Of interest here is whether a {\bf sustained} horizontal shear gives rise to sustained density layering and how the flow is organised in such a situation. Of course, it is important to remember that horizontal shear also has great relevance to oceanic flows,  where zonal jets provide horizontal shear but are also observed to develop vertical structure in the form of `stacked jets' \citep{Eden:2008hl,Hua:1997tk}. 

We force our model system via a sinusoidal body-forcing term, leading to what is  known as Kolmogorov flow \citep{Arnold:1960vx}, a flow known to support a rich array of spatiotemporal behaviour \citep{Lucas:2014ew}.  Stratified Kolmogorov flow has also been studied in two dimensions when the  shear is oriented vertically (\cite{Balmforth:2002,Balmforth:2005tf} who consider the linear instabilities, weakly nonlinear theory and direct numerical simulation of the flow). Here some layering connected to a stratified conductive instability is observed when the Prandtl number (i.e. the ratio of the diffusivity of the density field to the kinematic viscosity) is small. The three-dimensional extension of this work was recently discussed by \cite{Garaud:2015ce} who, motivated by astrophysical systems, consider the low P\'eclet number case. They investigate the limits of linear and energy stability and find that strong enough stratification will eventually suppress all instabilities. 

Our approach here is to make use of recent developments in (unstratified) shear flow transition where unstable exact coherent structures (ECSs) have been computed and shown to be responsible for organising the transition to and sustenance of turbulence \citep{Kawahara:2001ft,vanVeen:2006fm,Kerswell:2007ds,Viswanath:2007wc,2010PhST..142a4007C,Kreilos:2012bd,Kawahara:2012iu,2015arXiv150405825W}. This is made possible by employing a high dimensional Newton-GMRES-hookstep algorithm which is able to converge \emph{unstable} steady and time-periodic solutions efficiently from guesses taken from a chaotic simulation \citep{Chandler:2013fi,Lucas:2015gt}. This work is, to date and the authors' best knowledge, only the third work attempting to use such methods to understand the influence of stratification on a shear-driven flow. \cite{Olvera:2017ti} and \cite{Deguchi:2017dq} examine the effect of wall-normal stratification on certain known ECSs in plane Couette flow and find the states are heavily affected by stratification.  

The goal in this paper is to use this methodology to extract new solutions which exhibit the relevant coherent structure about which turbulence organises and forms layers in a stratified fluid. Such states are by definition nonlinear and their origins can be investigated by using continuation in parameter space to build a bifurcation diagram and determine if a robust connection can be made to linear instability mechanisms like the zig-zag instability. As such this study forms the first comprehensive investigation of layer formation in horizontally \emph{driven} simulations and the role of nonlinear ECSs in creating such layers. Such a simple basic flow and periodicity in all directions makes the system an efficient setting for the so-called `dynamical systems approach'. Fundamentally, our philosophy is to focus on `structures not statistics', as we aim to identify the characteristic, inherently nonlinear structures associated with (robust) layer formation in a stratified flow subjected to horizontal shear.

 The rest of the paper is organised as follows.
In section \ref{sec:form}, we present the formulation of the flow which we are considering.
In section \ref{sec:DNS}, we describe the results of our direct numerical simulations, where we do indeed observe the development of layered structures. In section \ref{sec:lin}
we present the results of a linear stability analysis of our flows, showing that there is a clear connection between the linear instabilities of the horizontally sheared stratified Kolmogorov
flow and the linear three-dimensional instability studied by \cite{DELONCLE:2007jl}. We then identify some
exact coherent structures in section \ref{sec:ecs} and demonstrate 
that they can be connected both to the observed zig-zag structures in the nonlinear simulations
and the predicted linear instabilities. We discuss our results and draw conclusions
in section \ref{sec:disc}.

%%%%%%%%%%%%%%%%%%%%%%%%%%%%%%%%%%%%%%%%%%%%%%%%%%%%%%%%%%%%%%%%%%%%%%%%

\section{Formulation}\label{sec:form}
We begin by considering the following version of the monochromatic body-forced, incompressible, Boussinesq equations

\begin{align}
\frac{\partial \bm u^*}{\partial t^*} + \bm u^*\cdot\nabla^*\bm u^*
 +\frac{1}{\rho_0}\nabla^*p^* &= \nu \Delta^* \bm u^* + \chi\sin(2\pi n y^*/L_y) \hx - \frac{\rho^*g}{\rho_0}\hz, \\ 
\frac{\partial \rho^*}{\partial t^*}+\bm u^*\cdot\nabla^*\rho^* +\bm u^*\cdot\nabla^*\rho_B&=  \kappa \Delta^* \rho^*\\
\nabla^*\cdot \bm u^* & =0 \\
\end{align}
where $\bm u^*(x,y,z,t) = u^* \hx+v^* \hy+w^* \hz$ is the three-dimensional
velocity field, $n$ is the forcing wavenumber, $\chi$ the forcing
amplitude, $\nu$ is the kinematic viscosity, $\kappa$ is the density diffusivity, $p^*$ is the pressure, $\rho_0$ is an appropriate reference density and $\rho^*(x,y,z,t)$ the varying part of the density away from the background linear density profile $\rho_B=-\beta z,$ i.e. $\rho_{\mathrm{total}}=\rho_0+\rho_B(z)+ \rho^*(x,y,z,t)$. Gravity acts in the negative $z-$direction. We impose periodic boundary conditions in all directions
$(x,y,z) \in [0,L_x]\times[0,L_y]\times[0,L_z]$ on $\bm u^*$ and $\rho^*.$

\begin{figure}
\begin{centering}
\includegraphics[width=0.6\textwidth]{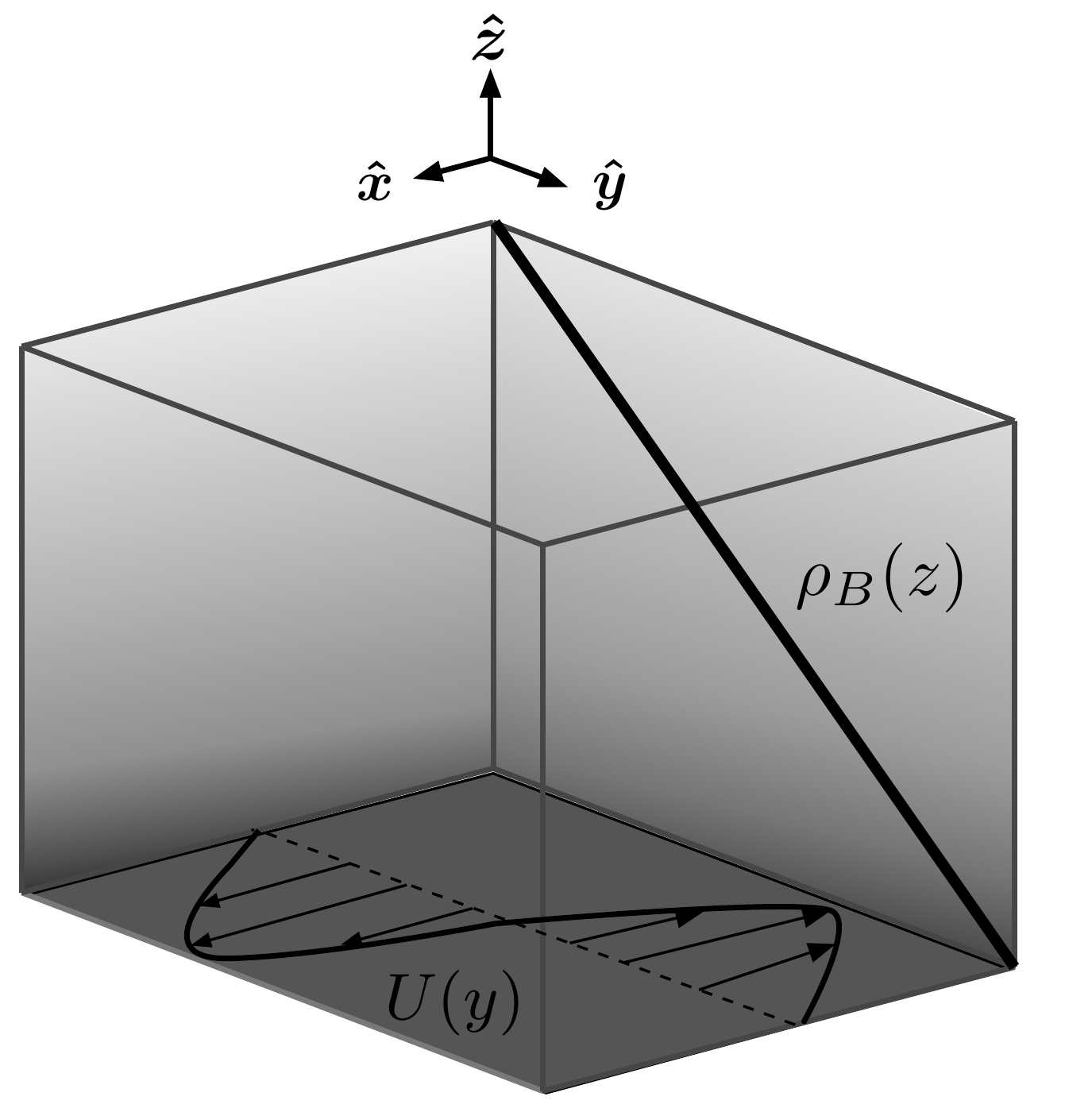}
\caption{\label{fig:scheme} Schematic showing the base flow and the background stratification.}
\end{centering}
\end{figure}

For simplicity we set $L_f=L_y=L_z$.
The system is naturally non-dimensionalised using the characteristic length scale $L_f/2\pi$, characteristic time scale $\sqrt{L_f/2\pi\chi}$ and density gradient scale $\beta=-\nabla \rho_B\cdot \hz$  to give
\begin{align}
\frac{\partial \bm u}{\partial t} + \bm u\cdot\nabla\bm u +\nabla p 
&= \frac{1}{Re} \Delta \bm u + \sin( n y)\hx - B \rho \hz\label{NSu},\\ 
\frac{\partial \rho}{\partial t} + \bm u\cdot\nabla\rho &= w + \frac{1}{Re Pr}\Delta \rho\\
\nabla\cdot \bm u &=0
\end{align}
where we define the Reynolds number $Re$, a buoyancy parameter $B$, the Prandtl number $Pr$ and 
the aspect ratio of the domain $\alpha$ as
\begin{equation}
Re := \frac{\sqrt{\chi}}{\nu}\left(\frac{L_y}{2\pi}\right)^{3/2}, \quad B:= \frac{g\beta L_y}{\rho_0 \chi 2 \pi}=\frac{N_B^2 L_y}{ 2 \pi \chi}=\frac{1}{4 \pi^2 F^2_{hB}}
, \qquad Pr := \frac{\nu}{\kappa}, \qquad \alpha = \frac{L_f}{L_x}. \label{eq:params}
\end{equation}
where $N_B$ is the (dimensional) buoyancy frequency associated with the background density field, 
and the characteristic velocity in  this particular (background)
definition of the horizontal Froude number  $F_{hB}=U/(L_f N_B)$
has been constructed using the characteristic length scale divided by the characteristic time scale,
exactly as in the definition of the Reynolds number.

It is important to note that the distinguished limit 
formally required of the scaling proposed by 
\cite{Billant:2001cs} 
is that the Reynolds number is large, 
(and so it is expected that the flow is turbulent)
while an appropriate horizontal Froude number $F_h$ is small such 
that $Re F_h^2$ is still large. For our flows, if we use this background definition of $F_{hB}$, 
these conditions  
correspond to  both $Re$ and (much more stringently) $Re/(4 \pi^2 B)$ being large. However, as discussed for example in \cite{Portwood2016}, there are
two significant, inter-related
issues which are of interest.
First, in practice, it is 
important to know the limits of applicability 
of a formally asymptotic scaling, identifying
the finite limits for the various parameters to be sufficiently
large or small so that the predicted regime actually arises.
Second, and related to this, there 
is of course a freedom (not least in terms of the selection of factors of $2 \pi$ for example)
in the specific form of the various characteristic 
length scales, and so comparison of specific delimiting numerical values of 
these non-dimensional parameters quoted in different
studies must be done with care.
As discussed in more detail by \cite{Portwood2016}, 
parameters based on the internal properties of any ensuing turbulent 
flow are more straightforward to compare from one flow to another (or indeed
from one sub-region of a flow to another) and so we  will also consider 
alternative measures for the key scaling  of the internal Froude number
in terms of the internal properties of the turbulence.

%$$N^2 = \frac{Ri2\pi^2\chi}{L_y^2}=RiRe^2 \nu^2\left(\frac{2\pi}{L_y}\right)^6$$
The equations then are solved over the cuboid $[0,2\pi/\alpha]\times[0,2\pi]^2$. We define the diagnostics involved in the energetic budgets as
\begin{align} 
\mathcal{K} &=\frac{1}{2}\langle|\bm u|^2\rangle_V, \qquad \mathcal{I} = \langle \bm u \cdot \bm f\rangle_V = \langle  u  \sin(ny) \rangle_V, \label{eq:diag1}\\
\mathcal{B} &= \langle \bm u \cdot B \rho\hz \rangle_V= \langle wB\rho \rangle, \quad
\mathcal{D} = \frac{1}{Re}\langle|\nabla\bm u|^2\rangle_V, \quad
\mathcal{D}_{lam} = \frac{Re}{2n^2},
\label{eq:diag2}
\end{align}
where 
$\mathcal{K}$ is the total kinetic energy density, $\mathcal{I}$ is  the energy input by the forcing,  $\mathcal{B}$ is the buoyancy flux, and $\mathcal{D}$ is the dissipation rate.  $\mathcal{D}_{lam}$ is the dissipation rate associated with the basic state
\beq \bm{u}_{lam} = \frac{Re}{n^2}\sin(ny)\hx. \label{eq:lam}\eeq
where the forcing and dissipation precisely balance and $\langle (\cdot)  \rangle_V:=\alpha \int \! \! \int \! \! \int (\cdot)\,dxdydz/(2\pi)^3$ denotes a volume average. We consider only $n=1$ throughout, i.e. the flow is forced with $\sin(y)\hx$ and denote a time average with an overbar, i.e.
$\bar{(\cdot )} = [\int_0^T (\cdot) \ud t]/T$ where T is normally the full simulation time {(having removed the initial 5 time units of transient spin-up from the initial condition)}.  
 Vorticity $\bm \omega = \nabla \times \bm u$ is used as the prognostic variable and direct numerical simulations (DNS) are performed using the fully dealiased (two-thirds rule) pseudospectral method with mixed fourth order Runge-Kutta and Crank-Nicolson timestepping implemented in CUDA to run on GPU cards. This code is a further extension, to stratified flow, of that used in \cite{Lucas:2017fz}. Each simulation is run on one GPU card and  due to memory limitations on the current generation of NVIDIA chips we are restricted to resolutions of $N_x=N_y=N_z=256.$  We initialise the flow velocity field's Fourier components with uniform amplitudes and randomised phases in the range $2.5 \leq |\bm k | \leq 9.5$ such that the total enstrophy $\langle |\bm \omega |^2\rangle_V = 1$ and leave the density field unperturbed initially i.e. $\rho=0$. Spatial convergence is checked by comparing the Kolmogorov microscale $\eta=\left(Re^3 \mathcal{D}\right)^{-1/4}$ with the smallest permitted scale in the system, i.e. the maximum wavenumber in simulation $k_{\mathrm{max}}=N_x/3=85,$ in other words we ensure $k_{\mathrm{max}}\eta >1.$

%%%%%%%%%%%%%%%%%%%%%%%%%%%%%%%%%%%%%%%%%%%%%%%%%%%%%%%%%%%%%%%%%%%%%%%%

\section{ Direct numerical simulations}\label{sec:DNS}
To determine the effect of stratification on this flow we begin by performing direct numerical simulations with various $B$ at fixed $Re=100,$ $Pr=1,$ $n=1,$ $\alpha=1$, integrating until $T=100.$  These calculations are designed to span a range of $B$ from the essentially unstratified to the point at which well formed layers are observed. They serve primarily as motivation for the stability and exact coherent structure analysis to follow. It is important to stress that the key aim of this paper is not to consider properties of flows in the LAST regime with extreme parameter values, but rather to explore the connections between linear instabilities, exact coherent structures and layer formation in a sheared and stratified fluid. 

\begin{figure}
%\begin{center}
\includegraphics[width=1.0\textwidth]{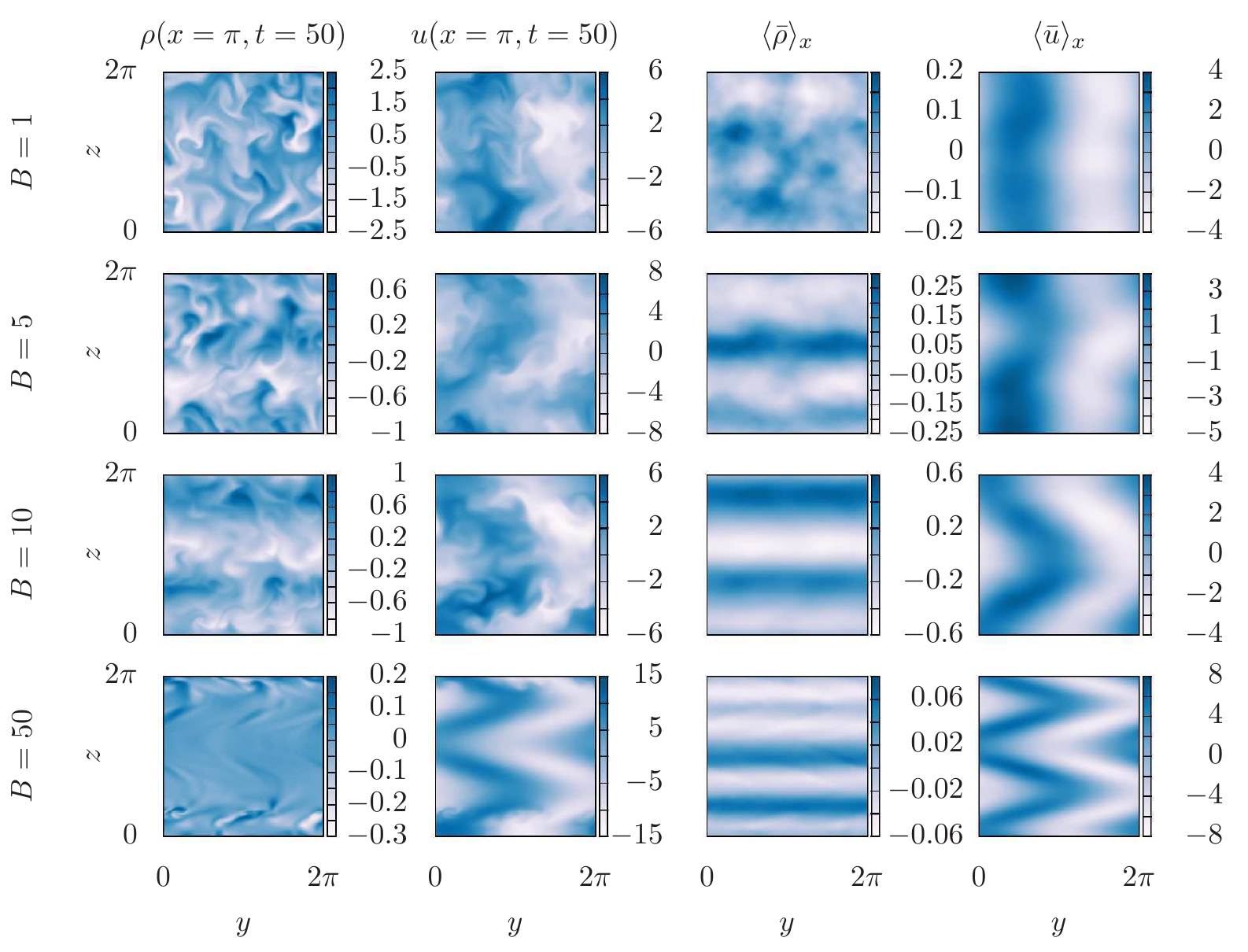}
%\end{center}
\caption{Snapshots in a $yz$ mid-plane (at $x=\pi$ for simulations with $Re=100$, $Pr=1$, $n=1$ and $\alpha=1$) of $\rho$ (first column) and $u$ (second column) at $t=50.$ The third column shows $\langle\bar{\rho}\rangle_x$, while the  fourth column shows $\langle\bar{u}\rangle_x,$ i.e. the $x-$average of the time-averaged (across the full $t\in [0,T]$ interval with $T=100$) perturbation density and streamwise velocity.  Rows are showing $B=1,5,10,50$ (simulations A1-A4 from table \ref{tab:DNS}) from top to bottom. Notice the increase in vertical structure as $B$ increases.\label{fig:DNS1}}
\end{figure}

Figure \ref{fig:DNS1} shows $yz$ planes (at $x=\pi$) of $t=50$ snapshots of $\rho$ and $u$ along with $\langle\bar{\rho}\rangle_x$ and $\langle\bar{u}\rangle_x,$ i.e. streamwise averages of the mean (in time over the whole simulation). Immediately we recognise the emergence of coherent structures organising the flow. Most evident in the mean, but also noticeable in the snapshots, is an inclining of the background horizontal shear into characteristic chevron, or (rotated) `v' shapes. Associated with this structure is the organisation of the density field into vertical layers, the number of layers (and associated v-shapes in $u$) increasing with $B.$ As the stratification is increased, the mean flow becomes progressively more layered. The formation mechanism of these layers, and in particular their relationship to both linear instabilities and exact coherent structures,  will be discussed in the subsequent sections.

 %We do comment here, however, that qualitatively the coherent structures emerging bear some resemblance to the early stages of the celebrated zig-zag instability of \cite{Billant:2000df,Billant:2000jg,Billant:2000wg}. In that case the base flow is a columnar vortex dipole, so that  the base states share some similarity (horizontally varying, vertically uniform). We will discuss in subsequent sections the connection with this instability. 

\begin{figure}
\begin{center}
%\scalebox{0.8}{\input{figs/throttle4}}
\includegraphics[width=1.0\textwidth]{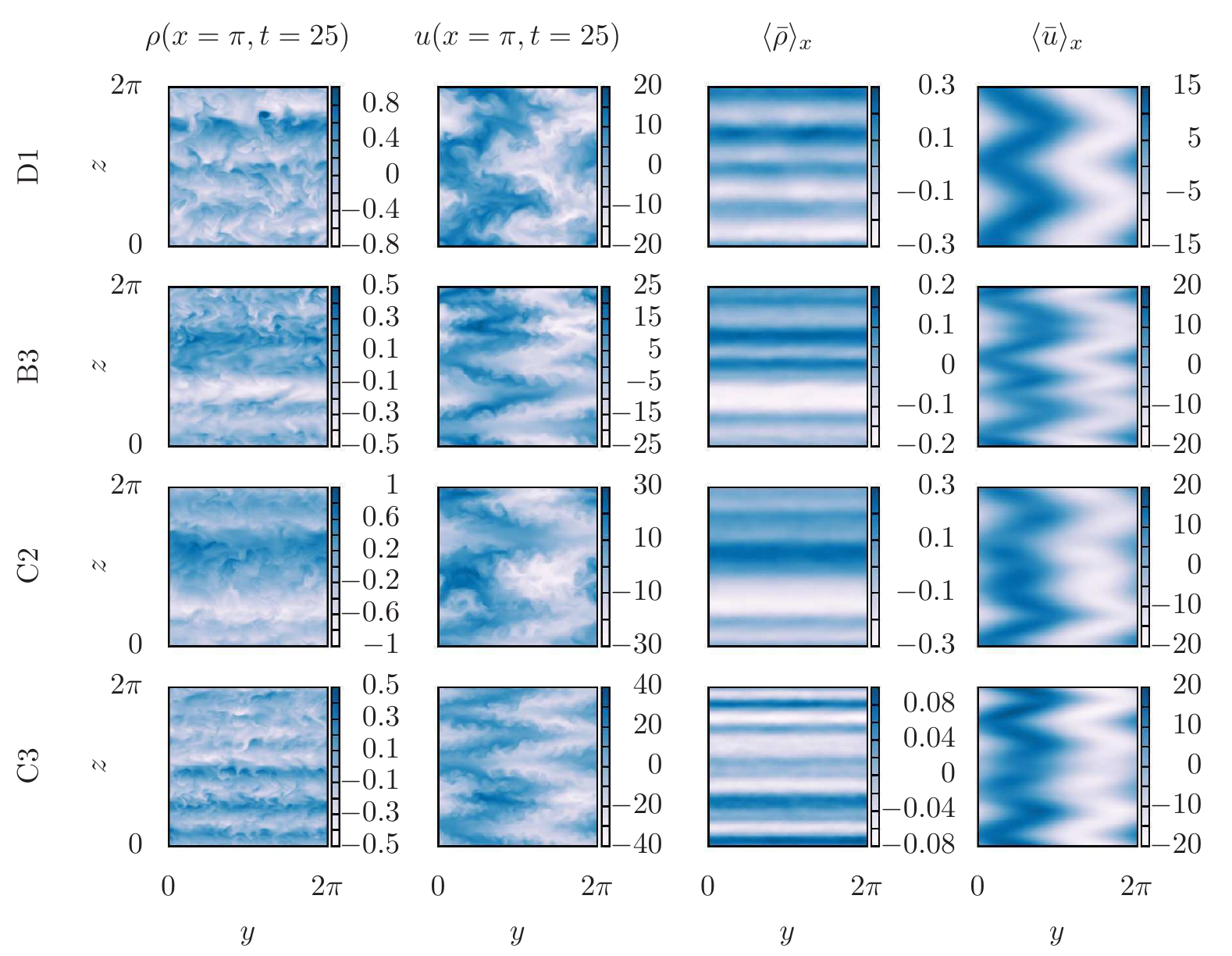}
\end{center}
\caption{Snapshots in a $yz$ mid-plane (at $x=\pi$) of $\rho$ (first column) and $u$ (second column) at $t=25$. The third column shows $\langle\bar{\rho}\rangle_x$ and the  fourth column shows $\langle\bar{u}\rangle_x,$ i.e. the $x-$average of the time-averaged perturbation density and streamwise velocity. Time averages are over $t\in[0,T]$ with $T=50$.  Rows are showing data from simulations D1, B3, C2 and C3 (as defined in table \ref{tab:DNS}) from top to bottom. Notice the increase in the number of layers as the buoyancy parameter $B$ increases and the increasingly angled structure of the velocity field. All simulations have $Pr=1.$\label{fig:throttle1}}
\end{figure}

\begin{figure}
\begin{center}
\includegraphics[width=0.95\textwidth]{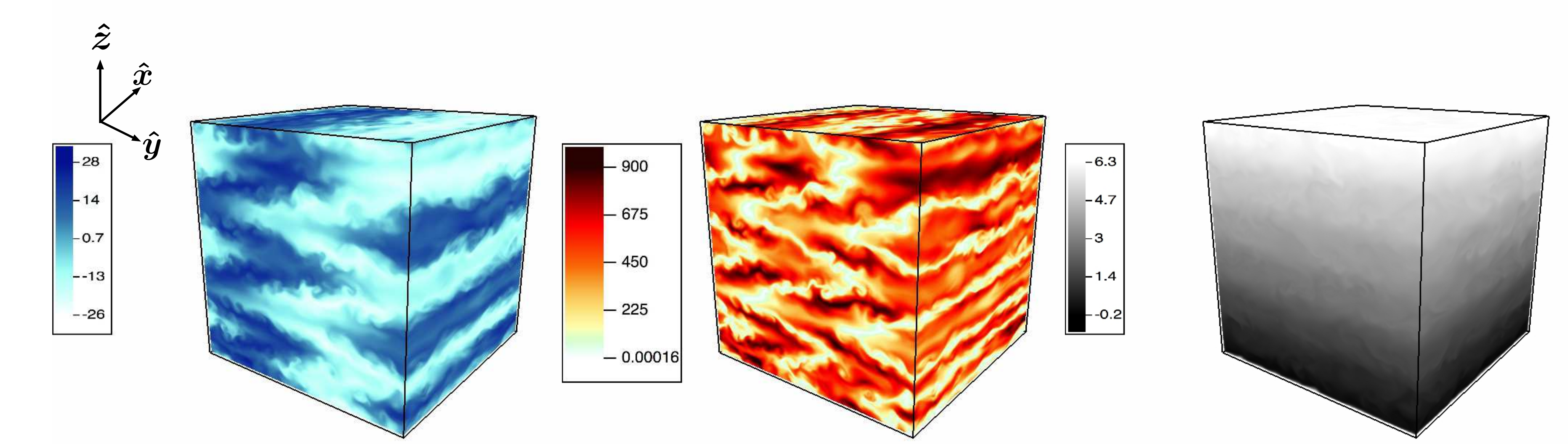}
\end{center}
\caption{Three-dimensional rendering of, from left to left right: the streamwise velocity $u$; the vertical shear $\frac{\partial u}{\partial z}$; and the total density field $\rho_B(z)+\rho,$ for the throttled simulation C2 with parameters as listed in table \ref{tab:DNS}. Note that regions of increased density gradient coincide once again with  quiescent regions in $u$ where the vertical shear is minimum. \label{fig:C2}}
\end{figure}

At $B\geq50$ the simulations exhibit significant bursts between quiescence and turbulence due to the combination of stratification suppressing vertical shear instability and body forcing across the domain. To obtain statistical stationarity we apply a throttling method to modulate the forcing strength to maintain turbulence and thus observe the formation of layers at increased stratification; without the throttling, bursting overturns the density field intermittently and does not permit well-defined layers to form. The results from large $B$ in the throttled cases are summarised in figures \ref{fig:throttle1} and \ref{fig:C2}. The important observation is the persistence, at large stratification, of the coherent chevron structures that are responsible for layer formation. Further details for the cases shown in figures  \ref{fig:throttle1} and \ref{fig:C2} are given in the appendix, as well as a detailed discussion of the bursting phenomena and our throttling protocol. 

%%%%%%%%%%%%%%%%%%%%%%%%%%%%%%%%%%%%%%%%%%%%%%%%%%%%%%%%%%%%%%%%%%%%%%%%

%%%%%%%%%%%%%%%%%%%%%%%%%%%%%%%%%%%%%%%%%%%%%%%%%%%%%%%%%%%%%%%%%%%%%%%%

\subsection{Layer length scale}

\begin{figure}
\hspace{-15mm}
\scalebox{0.55}{\input{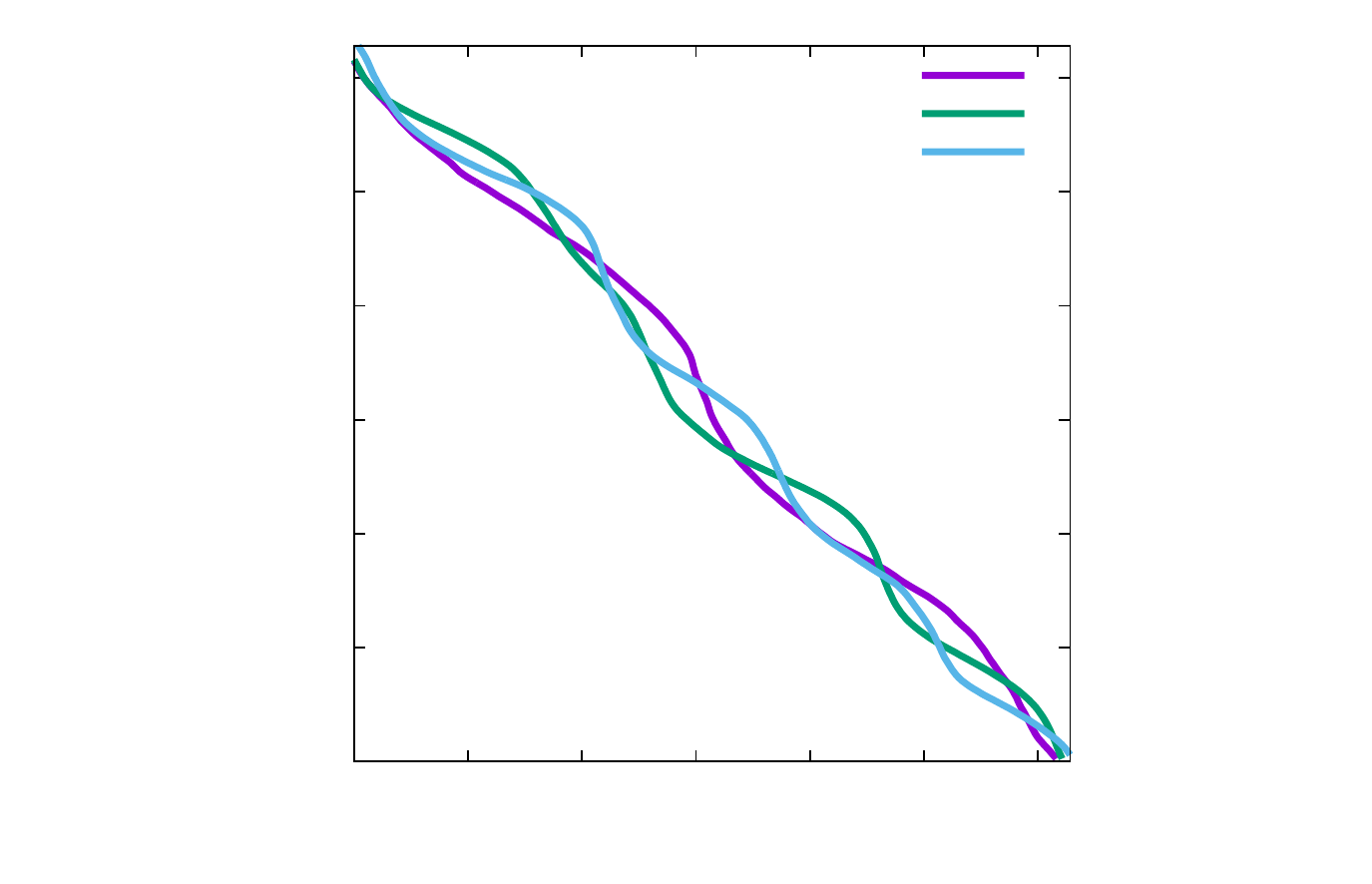}}
\hspace{-10mm}
\scalebox{0.55}{\input{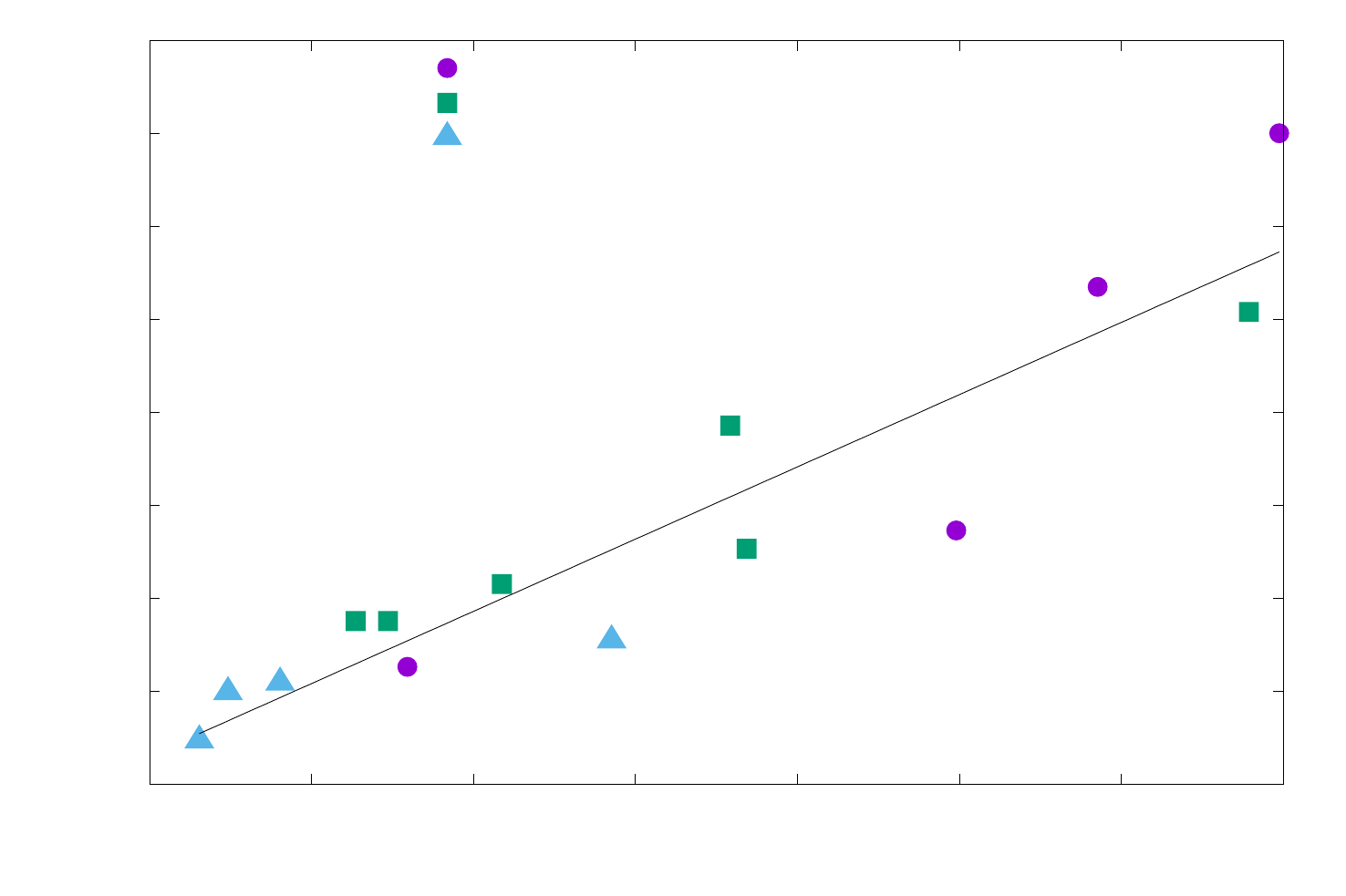}}
\caption{\label{fig:lu}Left: profiles of total density $\rho_{\mathrm{tot}}={\rho_B}+\rho$ at $t=50,$ $x=y=\pi$ for cases B1, B3 and C1 from table \ref{tab:DNS}. Right: Vertical lengthscale $l_{\bar{u}}$ 
as defined in (\ref{eq:ludef}) plotted against $U/N_B=\langle |\bar{(u)}| \rangle_{V}/\sqrt{B}$ for each DNS in table \ref{tab:DNS}. We find the scaling $l_v\sim U/N_B$ is recovered, particularly for smaller $U/N_B$. A least squares linear fit is also plotted: $0.78l_{\bar{u}} -0.12.$ }
\end{figure}

For the throttled simulations, it is now possible to observe smaller vertical length scales in the
developing layered structures. The increased buoyancy flux due to the sustained turbulence in the inclined shear layers now spontaneously leads to sharpening  interfaces separating relatively well-mixed `layers'. This spontaneous interface-layer formation can be most
clearly 
seen in figure \ref{fig:lu} (left panel) where profiles of the total density (including
the background linear component ${\rho_B}(z)$)
are plotted. Figure \ref{fig:C2} shows three-dimensional renderings of snapshots of the streamwise velocity, vertical shear $\frac{\partial u}{\partial z},$ and the total density field for the simulation C2 (see table \ref{tab:DNS}) with  $B=1000$ and $\mathcal{D}_0=100$. It can be
seen that gradients of density are approximately coincident with planes of small vertical shear of the underlying coherent structure, where the vertical buoyancy flux is 
small, or more precisely away from regions of strong vertical shear.

A common characterisation of vertical length scales in stratified turbulent flows is the scaling proposed by \cite{Billant:2001cs} leading to the `layered anisotropic stratified turbulence' (LAST) regime mentioned in the introduction. The characteristic
vertical
scale of the layers $l_v \sim U/N$ is commonly observed in simulation and experiments
 where $U$ is a typical velocity scale and $N$ is the appropriate buoyancy frequency. As mentioned in the 
introduction, this scaling is strictly
valid in a distinguished limit of 
small horizontal  Froude number, $F_h= U/l_h N\to 0,$
implying that $l_v \ll l_h$. 
Due not least to the periodicity of our computational domain, the formal
requirement that $l_v \ll l_h$ cannot be satisfied in our computations as the layers typically span the full horizontal extent of the box. However,
under the assumption that
the dissipation rate has an inertial scaling (see \cite{Brethouwer2007})  the LAST regime may equivalently be associated with sufficiently large
values of the buoyancy Reynolds number $Re_B$, defined as 
\begin{equation}
Re_B = \frac{\epsilon}{\nu N_B^2} = \frac{\mathcal{D} Re}{B}  \label{eq:rebdef}
\end{equation}
($\epsilon$ is the dimensional dissipation rate). As discussed in more detail in \cite{Portwood2016}, 
$Re_B$ is 
determined from the internal local characteristics
of the turbulence (captured by the dissipation rate) 
relative to the joint 
stabilizing effects of the fluid's viscosity and the 
background density gradient. 

Although it is now becoming widely accepted (see for example
\cite{Venayagamoorthy2016}) 
that $Re_B$ alone cannot
uniquely identify the properties 
of stratified turbulence driven by shear, the evidence is nevertheless
strong that $Re_B \gtrsim O(10)$ is 
necessary for there to be any opportunity for the LAST regime to occur.
As explained in detail by \cite{Brethouwer2007}, this 
is due to the necessity for there to be large scale separations between three
scales: the Kolmogorov microscale, $\eta=(Re^3 \mathcal{D} )^{-1/4}$;
the Ozmidov scale $l_O=(\mathcal{D}/B^{3/2} )^{1/2}$ and some vertical scale
of the layers  $l_v$, which must be smaller than the vertical extent of the computational domain. The Ozmidov scale is the largest vertical 
scale which is largely unaffected by the background stratification,
and so the ordering $\eta \ll  l_O < l_v $
ensures that there is both an inertial range for the turbulent motions
largely unaffected by the stratification, and 
a range of scales (between $l_O$ and $l_v$)
where the LAST regime occurs. Since by definition 
$Re_B = (l_O/\eta)^{4/3}$, $Re_B$ must be sufficiently large for an inertial range to exist. 

 The formal accessibility of this hierarchy of length scales  within our simulations is challenging. Due to the horizontal (periodic) boundary conditions, there is no natural way in which the actual {\bf horizontal} extent of any layer can be identified. Specifically it is not appropriate to use the characteristic length scale of the shear forcing as the horizontal scale of the `layer', as the flow has clearly adjusted so that the (density) layer extends horizontally across the entire computational domain. Furthermore, the 
observed vertical length scale
of layers must be sufficiently small so that it can be measured within the vertical extent of the computational domain, yet still sufficiently large so that it is both resolved computationally and also larger than
the   Ozmidov length scale $l_O$. In turn,
$l_O$ must still be sufficiently large
so that there is separation between it and the Kolmogorov microscale, $\eta,$ which, for the entire simulation to be resolved adequately, 
must be at worst of the same order as the smallest scale of the simulation.
These various length scales are listed in table \ref{tab:DNS}. It is 
important to appreciate that several of the simulations, particularly when $B$ is large, have values of $Re_B$ which are too small for the flow to be in the LAST regime. In particular such small values of $Re_B$ can arise when the dissipation rate $\mathcal{D}$ in the definition (\ref{eq:rebdef}) is taken to be the time-average of the actual (in general time-varying) volume-averaged dissipation rate within our simulations. That being said, the parameter choices for these DNS were purposely made to span situations from very weakly stratified cases, to cases where significant deformation of the mean flows are observed due to the stratification leading subsequently to the formation of layers.

It is therefore still of interest to investigate whether the 
 spontaneous layers observed here are
consistent with the scaling ideas at the heart of this proposed regime, even if the 
relevant parameters are not in the formally necessary self-consistent asymptotic scale separation.
We choose to 
define 
an appropriate
vertical `layer' length scale $l_{\bar{u}}$ for the layers
 as
\beq
l_{\bar{u}} = \frac{2\pi\int  \hat{\bar{u}} \ud k_z}{\int k_z \hat{\bar{u}} \ud k_z}, \label{eq:ludef}
\eeq 
where  $\hat{\bar{u}}$ is the Fourier transform
of the horizontally and  temporally averaged ($t\in[0,T]$) streamwise velocity $ \langle \bar{u} \rangle_{x,\,y}.$ 
For the LAST regime as discussed above, the layer length scale should scale with the ratio of an appropriate large scale or background  horizontal velocity scale $U$ and the (dimensionless) buoyancy frequency $N_B=\sqrt{B}$ from (\ref{eq:params}).
An appropriate choice for the velocity scale $U$ is the volume and time-averaged absolute streamwise 
velocity $U=\langle |\bar{u}| \rangle_{V}$.
Figure \ref{fig:lu} shows $l_{\bar{u}}$ plotted against $U/N_B$.  We find that for small $U/N_B$
 $l_{\bar{u}}$ approaches a linear trend, and  is much more scattered as $U/N_B$ reaches larger values, though still largely 
following  the asymptotic prediction of \cite{Billant:2001cs}, which is formally
expected to occur as $U/(L_h N_B) \rightarrow 0$. Significantly, the 
scaling appears to be satisfied even for the flows with relatively small buoyancy Reynolds number
(see table \ref{tab:DNS}) associated with relatively large values of $B$ and hence $N_B$, 
where the LAST regime scaling arguments are not strictly valid. 
As is shown from the least-squares fit, the {\bf vertical} Froude number $F_v$, 
defined as
\beq 
F_v = \frac{U}{N_B l_u} \simeq 0.78 = \mathcal{O} (1),\label{eq:fvdef}
\eeq
consistently with the scaling regime of \cite{Billant:2001cs}.

It should also be noted that similar estimates of layer depth could be constructed using the density field.
 However,  due to the strong spatiotemporal intermittency at large $B$ the trend to small $U/N_B$ is harder to observe in the density field, since  measurable layering of density requires sustained buoyancy fluxes and mixing, which are challenging to resolve.  At higher resolutions where turbulence can be maintained with larger target dissipation rates at larger $B$, it is perfectly reasonable to expect that robust  layering of the density field obeying the $U/N_B$ scaling
will be more straightforward to observe,  as is strongly suggested  in figure \ref{fig:C2}.  
{Note several other studies of stratified turbulence have observed this $U/N$ layer scaling, (e.g.  \cite{Holford:1999uc,2004JFM...517..281W, LINDBORG:2006eg,Brethouwer2007,Khani:2013,Oglethorpe:2013cv,Augier:2015fo,Thorpe:2016ga})}. However, 
 a key open question remains surrounding the precise mechanisms for the formation of such layers, which will be the focus of the subsequent sections. In particular, while a connection to known linear instabilities has been conjectured as the mechanism for such layers, {there} does not yet exist a robust connection between the output of (inherently nonlinear) direct numerical simulations and such an instability. 

%%%%%%%%%%%%%%%%%%%%%%%%%%%%%%%%%%%%%%%%%%%%%%%%%%%%%%%%%%%%%%%%%%%%%%%%

\section{Linear stability analysis}\label{sec:lin}

There appears to be at least 
qualitative similarity between the streamwise velocity structure  observed in our forced flows and the streamwise velocity eigenstructure  at onset of the now classic `zig-zag' instability mechanism of \cite{Billant:2000df,Billant:2000jg,Billant:2000wg}. Since this  class of  instability is  also often invoked as the precursor to layer formation (see e.g. \cite{Thorpe:2016ga}), 
it seems appropriate to investigate the linear stability properties of the flows
we are considering, in particular to identify whether they are prone to instabilities which may be identified as being of `zig-zag' type.
For the case when the forcing scale and streamwise integral scale are the same (i.e. using our notation $\alpha=1$ and $n=1$), the unstratified three-dimensional Kolmogorov flow is known to undergo subcritical transition to turbulence with the base flow remaining linearly stable at all $Re$ \citep{Marchioro:1986tg,Veen:2016fy}. Any linear instability in this geometry is therefore due to the added physics provided by the statically stable stratification.

The inviscid linear stability properties of a background horizontal $\tanh$ shear profile with linear background stratification was considered by \cite{DELONCLE:2007jl}. They  reported that, while two-dimensional
(in the horizontal plane)  homogeneous perturbations are the most unstable leading
to instabilities of KH type associated with the inflection point in the background shear profile, new inherently three-dimensional stratified instabilities are present and can have comparable growth rates. The growth rates of the stratified instabilities are also observed to follow the self-similar scaling with respect to an appropriate horizontal Froude number as proposed by \cite{Billant:2001cs}. 

\begin{figure}
\begin{centering}
\hspace{-5mm}
\scalebox{0.5}{\input{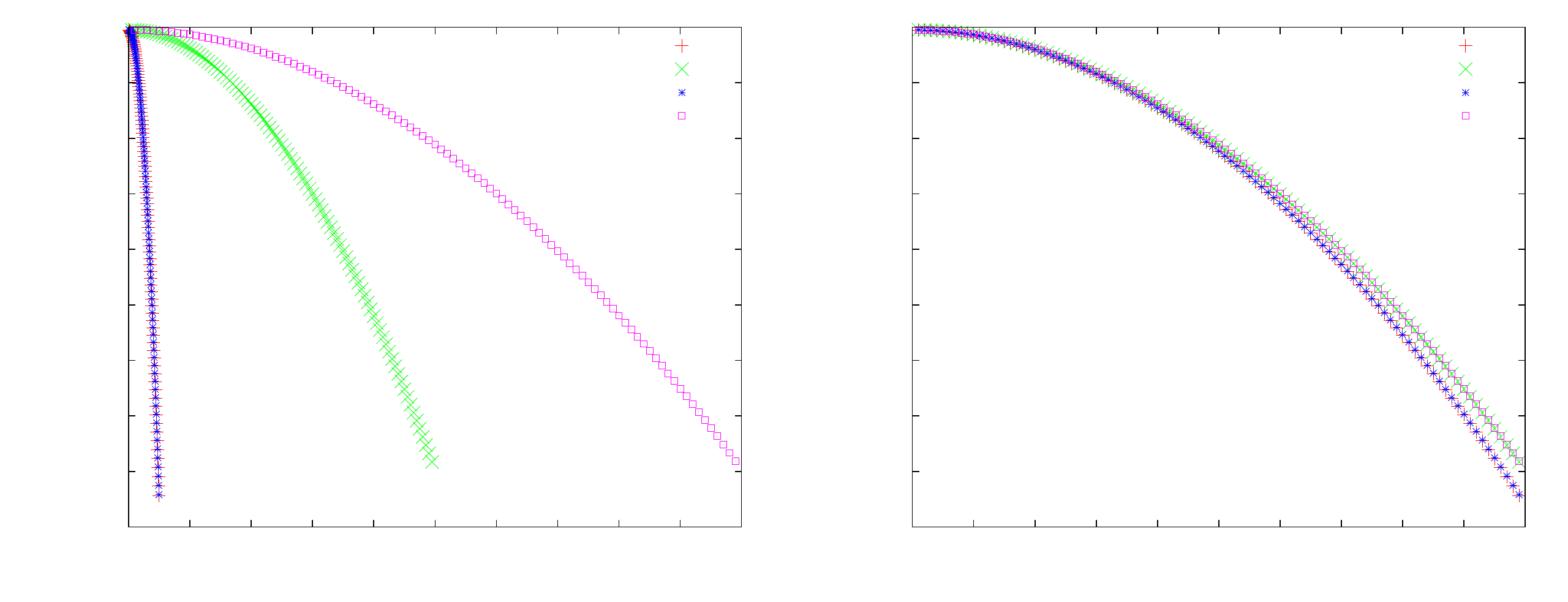}}
\end{centering}
\caption{\label{fig:grate}Variation of the  maximum (across all horizontal wavenumbers, associated
with $k_x=0.59$) and scaled growth rate $\sigma_m/Re$ with:   vertical wavenumber $k_z$ (left panel); and vertical
wave number scaled with stability horizontal Froude number (as defined in (\ref{eq:fhsdef})) $F_{hS}k_z$ (right panel).
 Red pluses mark results for a flow with  $Re=100,$ $B=10^4$ ($F_{hS}=1$), green crosses mark
results for  a flow with $Re=1000,$ $B=10^8$ ($F_{hS}=0.1$), red stars mark results for a flow with  $Re=1000,$ $B=10^6$ ($F_{hS}=1$) and magenta squares mark results for a flow with $Re=5000,$ $B=10^{10}$ ($F_{hS}=0.05$). }
\end{figure}

In the conventional fashion, we consider normal mode disturbances 
proportional to $\exp[ i ( k_x x + k_z z) + \sigma t]$,
away from the base state $S_B=(\bm{U}_B,\rho_B)$  
defined by
\beq
\bm{U}_B=\frac{Re}{n^2}\sin(ny)\hx,
\quad {\rho}_B=-z, \label{eq:sbdef}
\eeq
 and solve the ensuing eigenvalue problem for the linearised equations using the python NUMPY package, which is itself a front-end for LAPACK \citep{vanderWalt:2011dp}. We find very good agreement with the results of \cite{DELONCLE:2007jl} for the horizontal sinusoidal shear flow considered here provided $Re\gg 1$, which is 
a natural requirement as their analysis is inviscid. Figure \ref{fig:grate} shows the variation with vertical wavenumber  $k_z$ of the maximal growth rate  $\sigma_m$ (across all streamwise wavenumbers, corresponding to $k_x=0.59$)  rescaled with $Re$ (since $\sigma_m \sim Re$), plotted  
for a range of $Re$ and $B.$ 
Similarly to the results of  \cite{DELONCLE:2007jl}, we also find that two-dimensional disturbances (with $k_z=0$) are most unstable and that the most unstable horizontal wavenumber $k_x$  is independent of $B.$  This is unsurprising, as this two-dimensional
instability is once again of KH type associated with the inflection 
point in the background velocity shear.
For further comparison with their results, it is necessary to define an appropriate horizontal Froude number for this stability problem. 
An appropriate velocity scale for comparison for these linear instabilities is the maximum magnitude  of $\bm{U}_B$ i.e.
$Re/n^2$, (half the total velocity jump across the shear layer) while an appropriate length scale is (within our nondimensionalization) $1/n$, and so the stability 
horizontal Froude number $F_{hS}$ may be defined as
\beq
F_{hS}=\frac{Re}{n\sqrt{B}}. \label{eq:fhsdef}
\eeq

As already noted, for  all the calculations we present here, we set $n=1$. Replotting the growth rates against $F_{hS}k_z$ leads to  a very good collapse of the curves for a wide range of stability horizontal Froude numbers $F_{hS}=(0.05, 0.1, 1)$ as shown  in the right panel of  figure \ref{fig:grate}, 
once again in pleasing agreement with \cite{Billant:2001cs} and \cite{DELONCLE:2007jl}. In particular,  see figure 4 in \cite{DELONCLE:2007jl} where an equivalent  definition of horizontal Froude number (i.e. using 
the maximum magnitude of the velocity distribution  and the scale of the shear layer) are used and the curves are extremely similar to those in figure \ref{fig:grate}; the shape of the shear profile serves only to shift the growth rates slightly. We stress again that this is the appropriate Froude number for comparison of linear stability results, and not as the characteristic horizontal $F_h$ of the layers which develop in the DNS, as they extend horizontally across the entire computational domain.

\begin{figure}
\begin{centering}
\scalebox{0.75}{\input{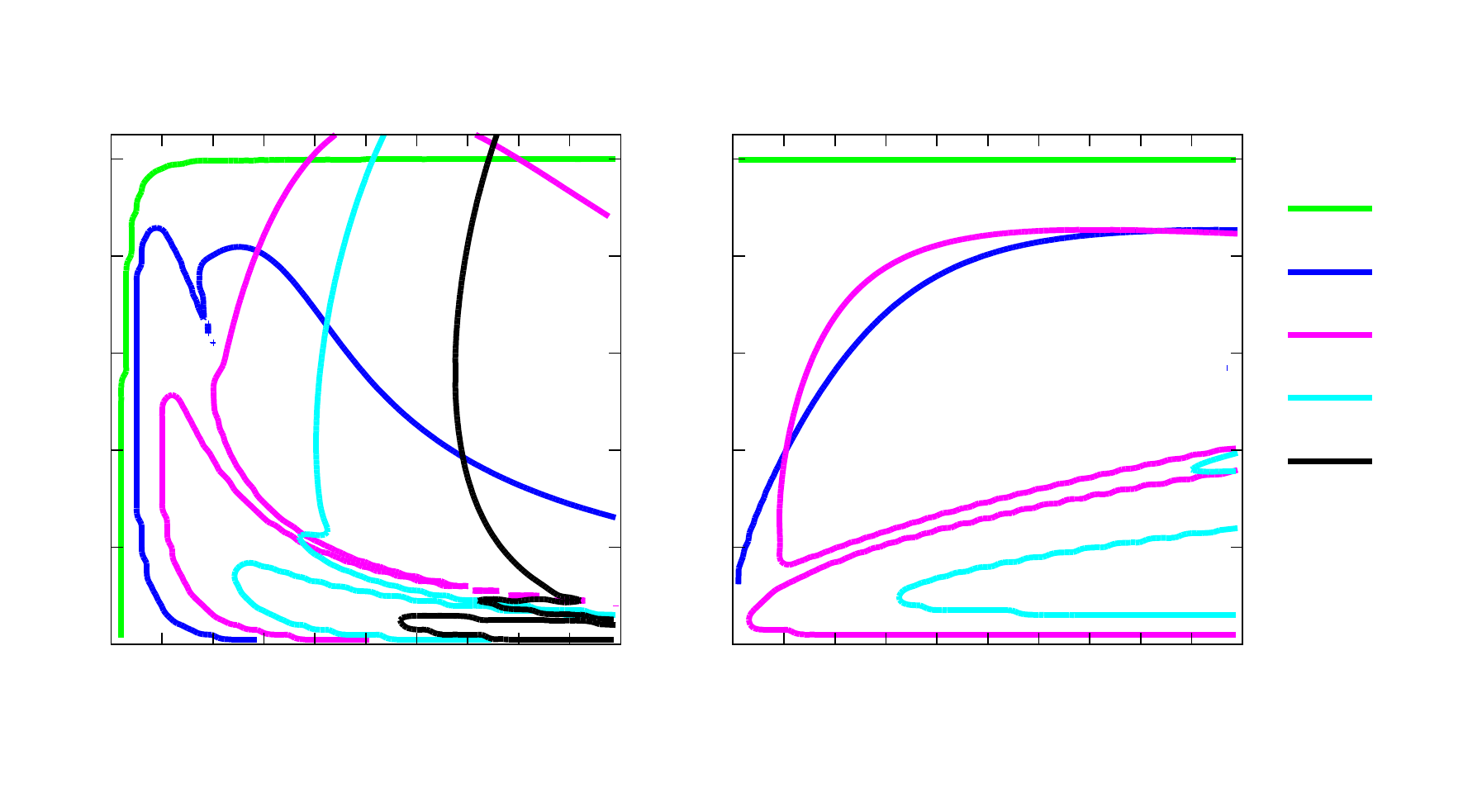}}
%\parbox[b]{1.2\textwidth}{\hspace{-25mm}
%\scalebox{0.7}{\input{figs/stab_Re}} \hspace{-40mm}
%\scalebox{0.7}{\input{figs/stab_B}}
%}
\end{centering}
\caption{\label{fig:stab} 
Left panel: Neutral stability curves on the $Re-k_x$ plane
for modes with vertical wavenumbers $k_z=0,1,2,3,4$ in a flow with $B=50$;
Right panel:  
Neutral stability curves on the $B-k_x$ plane
for modes with vertical wavenumbers $k_z=0,1,2,3,4$ in a flow with $Re=15$.
Note that the $k_z=0$ instability has $k_x < 1$ for all $Re$ and $B$.}
\end{figure}

In the (nonlinear) DNS,    only horizontal wavenumbers  $k_x=\frac{m}{\alpha}$ (where $m$ is an integer)  are admissible. Therefore, any linear instabilities must have $k_x \geq 1/\alpha$ to have any chance of occuring within our computational domain.
In figure \ref{fig:stab} (left panel) for flows with $B=50$ we plot neutral curves of linear stability for various vertical wavenumbers $k_z$ on the $Re-k_x$ plane. Although the unstratified long-wave instability is clearly evident for $k_z=0$, the neutral curve asymptotes near $k_x=1$ but crucially never crosses as $Re$ increases. Similarly,  the neutral curve for $k_z=1$  always  has $k_x < 1,$ i.e. a long streamwise wavelength is required in this range of $Re$, 
and so neither the instabilities with $k_z=0$ nor $k_z=1$  are expected to arise in the direct numerical 
simulations discussed above, with aspect ratio $\alpha=1$. 
 However,  the neutral curves for the instabilities with $k_z>1$  do intersect $k_z=1$ (and indeed cross to even higher 
wavenumbers), suggesting that  these instabilities can actually develop within our  direct numerical simulations where $\alpha=1$. The right plot in figure \ref{fig:stab} shows the neutral curves on a $B-k_x$ plane,  for fixed $Re=15$,  illustrating that
increasing $B$ and $Re$ introduces more  instabilities with higher values of vertical wavenumber $k_z$. 

%%%%%%%%%%%%%%%%%%%%%%%%%%%%%%%%%%%%%%%%%%%%%%%%%%%%%%%%%%%%%%%%%%%%%%%%

\section{Exact coherent structures}\label{sec:ecs}

%While the coherent patterns observed in the DNS (i.e. figure \ref{fig:DNS1}) do not appear to have the structure of a normal-mode instability of $u_{lam}$, 
We can determine if the finite-amplitude states observed in the direct numerical simulations can be connected to the inherently stratified linear instabilities discussed in the previous section by attempting to converge invariant states using a Newton-GMRES-hookstep algorithm \citep{Viswanath:2007wc}. The approach used here is simply to sample the time series of a certain DNS which exhibits a close approach to the coherent structure of interest and output the full state vector for post-processing. The state vector in this case consists of the full flow field and density degrees of freedom;
$$ \bm{X}= \begin{bmatrix} u \\ v\\w\\ \rho \end{bmatrix}$$
 The post-processing takes the form of a high dimensional root-finding algorithm which solves
$$\bm{F}(\bm{X_0},T_p):=\bm{X}(\bm{X}_0,T_p)-\bm{X}_0 =0,$$
where $\bm{X}_0$ is the starting condition and $\bm{X}(\bm{X}_0,T_p)$ is the final state after some small, fixed period trajectory of length $t=T_p.$ This period $T_p$ is arbitrary in the cases considered here since searches are only conducted for steady states. Note that the final state vector is a highly nonlinear function of the starting state $\bm{X}_0.$ Each Newton iteration updates $\bm{X}_0^{new}=\bm{X}_0+\delta\bm{X}_0$ where $\delta\bm{X}_0$ is given by 
$$ \frac{\partial\bm{F}}{\partial\bm{X_0}}\delta\bm{X_0} = -\bm{F}(\bm{X_0},T_p)$$
The Newton-GMRES-hookstep method then forms the solution to the ensuing linear system of derivatives $\partial\bm{F}/\partial\bm{X_0}$ via GMRES, a Krylov subspace method and a simple forward difference to approximate the action of the matrix. A hookstep then constrains each Newton step to a trust region inside which the Newton linearisation is expected to hold. The method used was first described by \cite{Viswanath:2007wc} and the implementation here is identical to that described in detail in \cite{Chandler:2013fi}, except for the inclusion of the GPU Boussinesq timestepping code and additional degrees of freedom for density. Note in general we search for translations in the solution (as explained in \cite{Chandler:2013fi}) due to the continuous symmetries in $x$ and $z$ which allow travelling waves to be converged. However, all the states under discussion here are stationary. 
Successfully converging invariant states and performing arc-length continuation of the solutions in a certain control parameter (for example $\alpha,$ $Re,$ or $B$) enables us to build a bifurcation diagram for these invariant  states. This continuation method includes the relevant parameter into the state vector $\bm{X}$ as an unknown and the solution is extrapolated along the solution branch via a local arc-length parameterisation which permits the solution curves to turn corners.

\begin{figure}
\begin{centering}
\scalebox{0.9}{\input{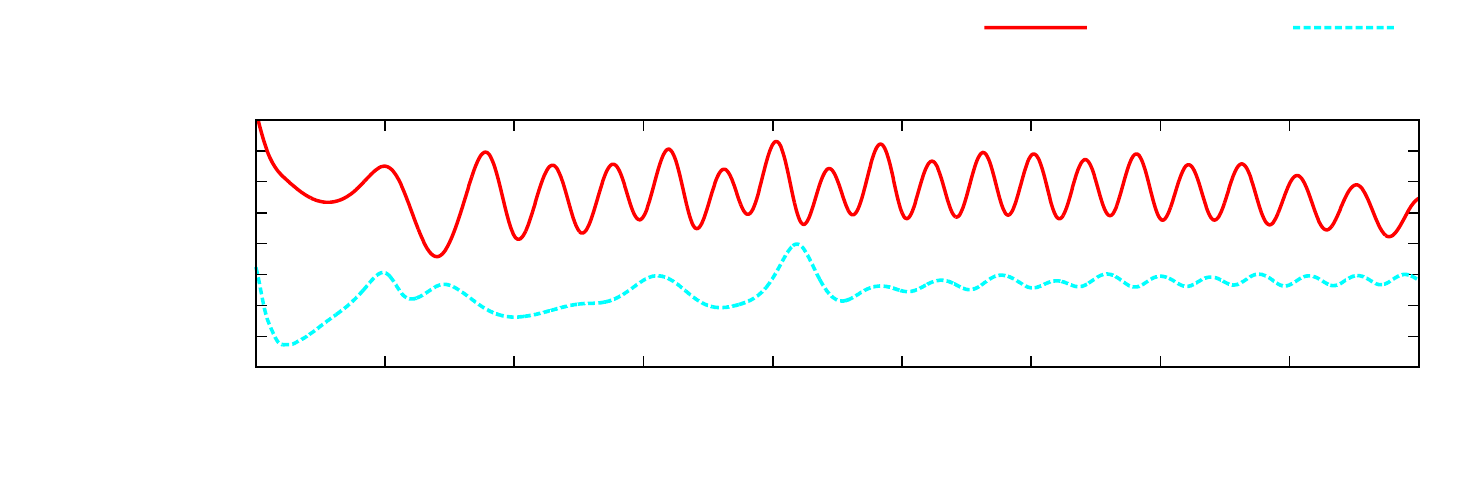}}
\end{centering}
\caption{Time series of $\mathcal{D}/\mathcal{D}_{lam}$ for the cases $Re=8$ and $Re=15$ for $B=50$ and $\alpha=0.9$\label{fig:D_Apt9}.}
\end{figure}

In practice, the greatest difficulty in this case (and in general) is in finding a starting $\bm{X}_0$ sufficiently nearby in phase space to an underlying unstable steady solution for the Newton method to converge. For this reason we concentrated on unthrottled flows with low $Re$ in relatively long domains with $\alpha<1$ when attempting to identify invariant states. In such situations,  the flow is less `unstable', in the specific sense that relatively simple coherent structures are observed to be approached closely. Therefore we expect our  attempts at convergence to meet with more success. A set of new simulations were performed at a range of low $Re$ and at various values of $B$ and state vectors were sampled and fed into the Newton-GMRES method. Our successes at $Re=8$ and 15 at $B=50$ and $\alpha=0.9$ are reported here. 
Figure \ref{fig:D_Apt9} shows a time series of the dissipation for these cases to highlight that, despite the low $Re,$ these flows are unsteady and, at least for these short times, they exhibit chaotic dynamics. It is quite possible the long time attractor is periodic or quasi-periodic, particularly at $Re=8.$ This issue is not of immediate interest as we use these chaotic trajectories to sample for a nearby unstable steady solution. We observe the closest approach to the coherent structure when $\mathcal{D}/\mathcal{D}_{lam}$ has a minimum and so use this as a guess for the Newton-GMRES algorithm. 

\begin{figure}
\begin{centering}
\includegraphics[width=\textwidth]{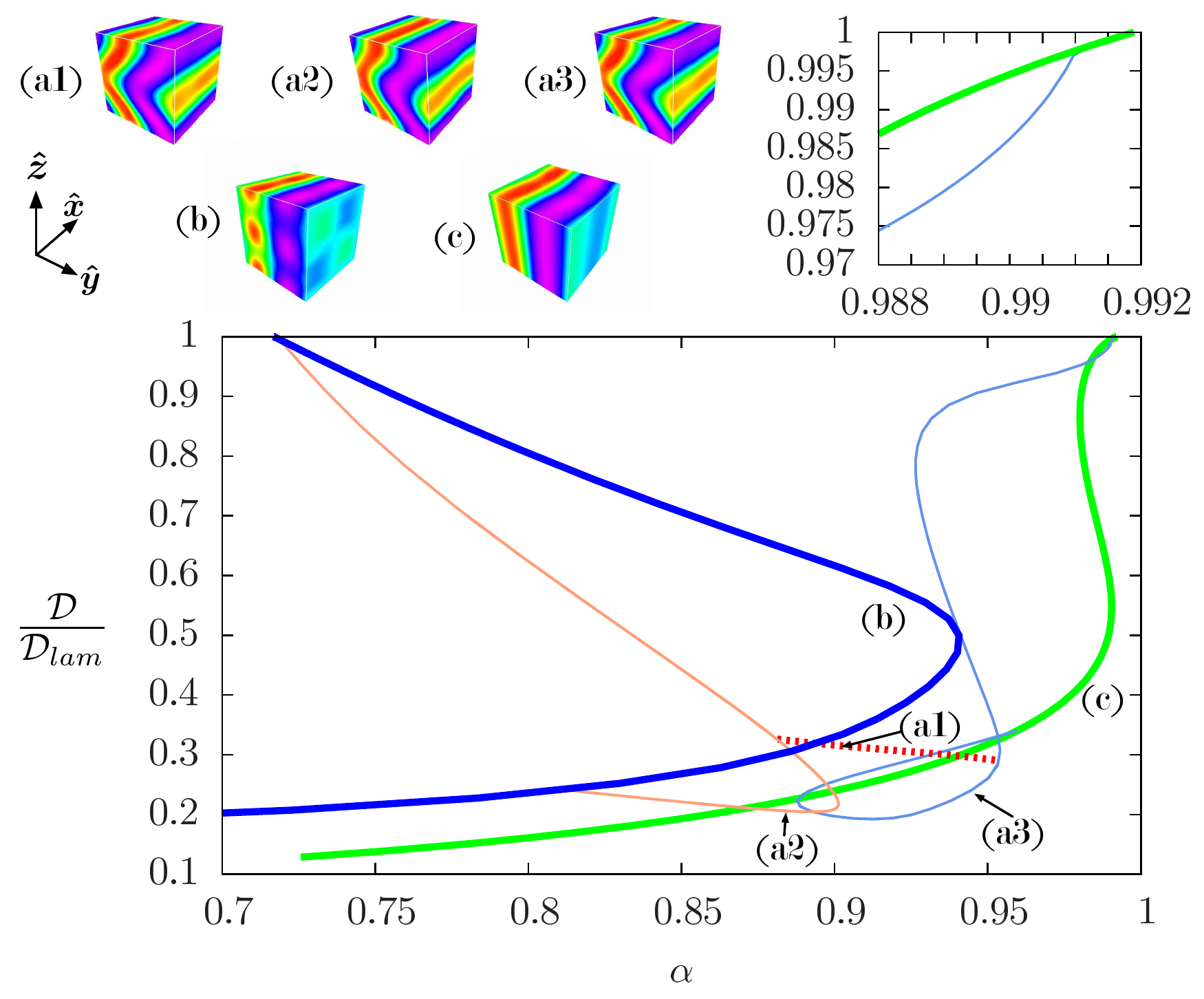}
\end{centering}
\caption{\label{fig:BIF} Bifurcation diagram showing solution branches for flows
with $Re=8$ and $B=50$ identified through continuation in $\alpha$ (or equivalently streamwise wavenumber $k_x$) against scaled total dissipation $D/D_{lam}.$ The spatial structures of streamwise velocity for states (a1)-(c) are shown in the top left. Thick lines correspond to primary states generated by linear instabilities of the laminar velocity profile $u_{lam},$ (equation (\ref{eq:lam})), thin lines correspond to secondary states generated by linear instabilities of the primary states and dashed lines correspond to tertiary states, in turn arising from linear instabilities of the secondary states. The green curve follows  the two-dimensional solution state  (c) (with $k_z=0$) connecting to the long-wave homogeneous instability of the laminar base flow. The inset shows a zoom of the bifurcation leading to state (c) and also shows the (secondary) stratified instability from this state leading  to the three-dimensional inclined `V'-shaped coherent structure (a3) which follows  the lighter blue curve as $\alpha$ is continued, 
and which then through a further tertiary bifurcation (marked with a dashed red line)  switches to state (a1).
State (a1) follows the red dashed curve in this plane, and in turn attaches to the secondary state (a2) plotted with a thin orange line. State (a2) finally connects to the primary state (b), plotted with a thick dark blue line,  which originates from the  $k_z=1$ stratified linear instability of the laminar base flow $u_{lam}$. Note that the labelling of the curves corresponds to the locations in parameter space for the three-dimensional visualisations plotted above.}

\end{figure}

Figure \ref{fig:BIF} shows the results arising from converging the unstable invariant state (labelled state (a1)) in a flow with $Re=8,$ $B=50$ and $\alpha=0.9$.   
This  state clearly  has one chevron-shaped structure in its streamwise velocity as shown in the top left panel. 
Due to the assumed periodicity in the vertical direction, this state will naturally form a `zig-zag', as clearly observed in the experiments of \cite{Billant:2000df}.  

%It is important to remember that this state is inherently nonlinear, and that we have been able not only to identify its modal structure, which undoubtedly has close connection with the eigenstructure of the linear instability, but crucially also its saturated amplitude, demonstrating unequivocally that these instabilities are indeed of `zig-zag' type.
The lower panel of figure \ref{fig:BIF} shows the bifurcation diagram mapped out by continuing this solution in $\alpha$, projected onto $\mathcal{D}/\mathcal{D}_{lam}$ such that when this quantity is identically 1, the solution has returned to $u_{lam}$ defined in equation (\ref{eq:lam}).  Solution (a1) follows the (dashed) red line and is observed to connect, via two bifurcations (and via state (a2)) to the state (b) which is the ensuing state arising from the $k_z=1$ linear instability. Specifically, as here the streamwise dependence of the steady states is a domain scale long wavelength, the $\alpha$ of the continuation and the $k_x$ of the linear instability mode in figure \ref{fig:stab} are equivalent as $\mathcal{D}/\mathcal{D}_{lam} \to 1.$ In other words, the bifurcation of the \emph{nonlinear} steady state (b) from $u_{lam}$ at $\alpha\approx0.717$ corresponds directly with crossing the $k_z=1$ neutral curve in figure \ref{fig:stab} at $Re=8$ and $B=50$ and $k_x\approx 0.717$ from the linear analysis. 
Since  state (b) arises from a linear instability of the laminar flow, it is marked with a thick line (blue for consistency with the $k_z=1$ neutral curves in figure \ref{fig:stab}) and we
refer to it as a `primary' state, while we refer to 
state (a2) as a `secondary' state and plot its continuation with a thin line, arising as it does from a
bifurcation from a primary state. We call (a1) a `tertiary state' since it bifurcates from a secondary state and denote it with a dashed line.  Note, consistently with the linear stability calculations presented in figure \ref{fig:stab}, state (b) does not exist in a flow geometry with $\alpha=1$, but requires
$\alpha < 1$. 

Moreover when increasing $\alpha$, state (a1) connects via a bifurcation 
to another secondary state (a3), marked with a thin blue line. State (a3) is found to connect to the primary state, labelled (c), and marked with a thick green line (again for consistency with figure \ref{fig:stab}) arising from the $k_z=0$ unstratified linear instability of the laminar flow. As before, the bifurcation at $\alpha\approx0.992$ is associated with the corresponding neutral curve for $k_z=0,$ at $Re=8,$ $B=50$ and $k_x\approx0.992.$ In other words, the state labelled (c) undergoes its own stratified linear instability (to the state (a3) marked with the light blue line) in much the same fashion as the laminar background flow $u_{lam}$ undergoes the linear instability, ultimately leading to the saturated state (b). The different varieties of chevron shaped states (a1)-(a3) are very similar and only on close inspection is one able to discern the subtle symmetry differences, notably between (a2) and (a3) which are switched by traversing the `rung' of the tertiary state (a1). Note that  all states are inherently three-dimensional apart from the states lying along the $k_z=0$ green curve associated with state (c). In summary, we have identified an inherently nonlinear and coherent chevron-shaped structure which is manifest in the states labelled (a). Furthermore these states are highly reminiscent of the finite amplitude structures observed  experimentally by  Billant and Chomaz. Crucially, we establish that the chevron-shaped structure that we have isolated arises from a {\bf secondary} bifurcation away from the nonlinear state associated with the primary instability, and should not be thought of 
as the straightforward finite amplitude manifestation of the 
primary instability at least for this particular instability with $k_z=1$ within our flow domain. This is consistent with the numerically-based observations of \cite{2008JFM...599..229D} that the zig-zag instability does not `saturate'. To be specific, the nonlinear state labelled (b) has a translational invariance $(x,y,z) \rightarrow (x+\pi/\alpha,y,z+\pi)$ and this symmetry is broken by the bifurcation to the secondary state (a2). 

\begin{figure}
\begin{centering}
\includegraphics[width=\textwidth]{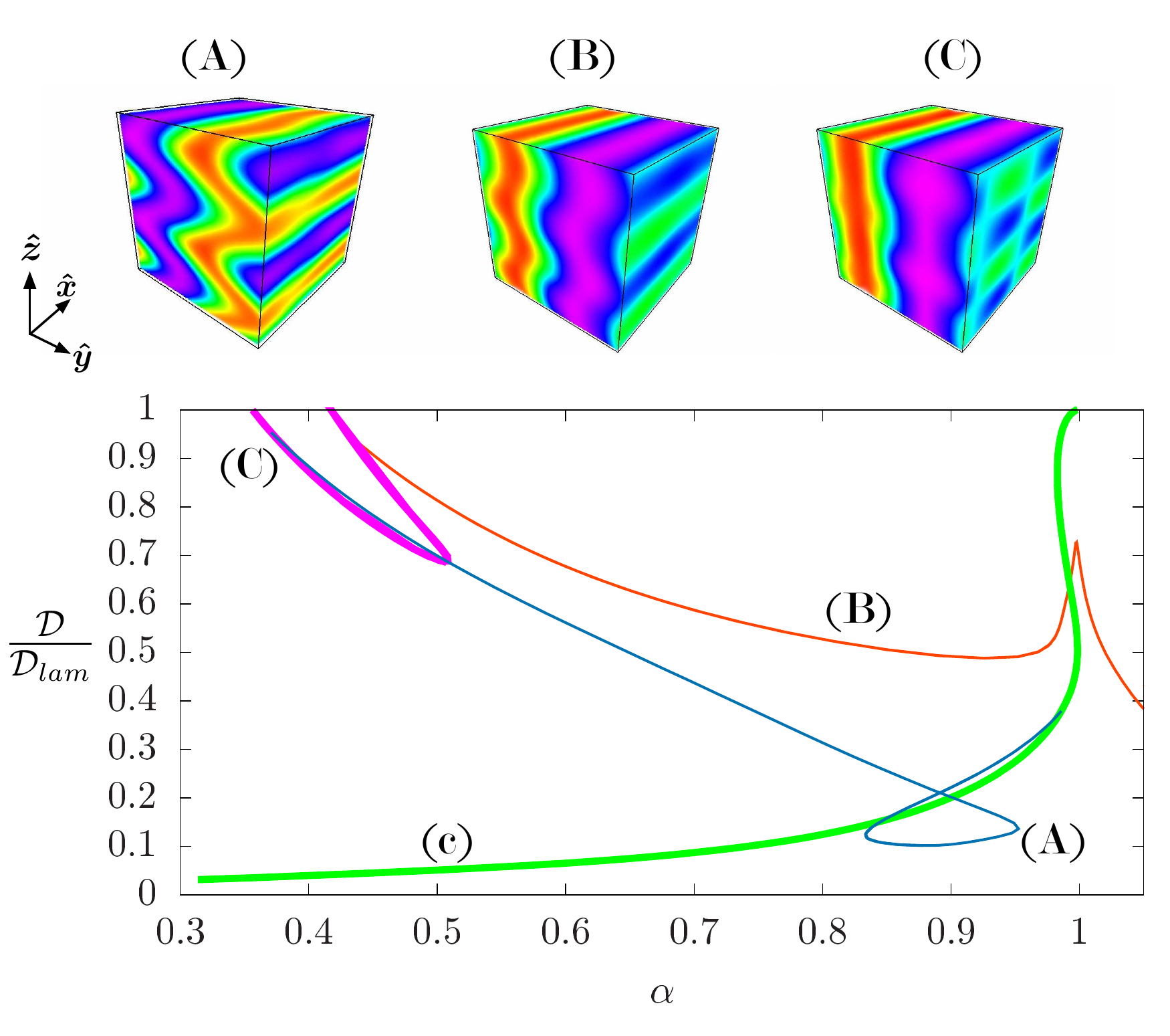}
\end{centering}
\caption{\label{fig:BIF2} Bifurcation diagram showing solution state branches 
for flows
with $Re=15$ and $B=50$ identified through  continuation in $\alpha$ (or equivalently streamwise wavenumber $k_x$) against scaled total dissipation $D/D_{lam}.$
The spatial structures of streamwise velocity for states (A)-(C) are shown in the upper panels.
As in figure \ref{fig:BIF} the thick green curve shows the evolution of the  primary state (c) (now at $Re=15$ not 8) arising from the linear instability with $k_z=0$, 
while the  thick magenta curve shows the evolution of primary state (C), arising from the $k_z=2$ linear instability of the laminar base flow $u_{lam}$. The thin orange line shows the evolution of the  secondary state (B) 
and the thin blue line shows the evolution of the secondary state (A), both generated by instabilities of the primary state (C). State (A) also attaches back to the primary state (c) with $k_z=0$. Labels are placed adjacent to where the three-dimensional visualisations are made in parameter space.}
\end{figure}

The states shown in figure \ref{fig:BIF} ultimately arise from the $k_z=0$ and $k_z=1$ primary 
instabilities of the laminar base flow, and so are associated with flow geometries with relatively long domains with $\alpha$ strictly less than one  (though in some cases quite close to one). However, this is not a necessary restriction. 
Figure \ref{fig:BIF2} shows the analogous situation for states
bifurcating from the $k_z=0$ linear instability (once again plotted with a thick green line for consistency with the neutral curves plotted in figure \ref{fig:stab}) and from  the  $k_z=2$ linear instability (plotted with a magenta line,
also consistently with figure \ref{fig:stab}). Here,  the nonlinear coherent structure is initially found from a guess obtained from the $Re=15,$ $B=50$ and $\alpha=0.9$ case.  
The state (denoted (A))  has two chevron-shaped inclined shear layers, as shown in the upper leftmost panel of figure \ref{fig:BIF2}, and once again is reminiscent of finite amplitude `zig-zag' structures. 

Perhaps unsurprisingly, through continuation (plotted with a thin blue line in the figure) this state is once again found to be `secondary', in that it attaches through bifurcations both to the primary state (shown
in the upper righthand panel of figure \ref{fig:BIF2} as state (C), and plotted with 
a thick magenta line for consistency with figure \ref{fig:stab})  arising from the linear instability with $k_z=2,$
and the primary state arising from  the two-dimensional ($k_z=0$) linear instability, whose continuation
with respect to aspect ratio $\alpha$ is once again plotted with a thick green line. Note that analogously to the $k_z=1$ case discussed in figure \ref{fig:BIF}, the secondary state (A) breaks the $(x,y,z) \rightarrow (x+\pi/\alpha,y,z+\pi/2)$  translational symmetry of primary state (C). 

Using continuation, we also identify 
another secondary state, labelled (B) and plotted with a thin orange line. This state is,  in some sense, lower amplitude, having $\mathcal{D}/\mathcal{D}_{lam}$ nearer to 1,  and does not have such strongly inclined layers, although a chevron-shaped structure is still apparent in the streamwise velocity, as shown in the middle upper panel of figure \ref{fig:BIF2}. Importantly however, we find that state (B) still persists to $\alpha\geq 1.$ Notably, the local maximum in the solution curve is found at $\alpha\approx 0.997$ and forms a smooth maximum, despite looking like a cusp on this $(\alpha, \mathcal{D}/\mathcal{D}_{lam})$ projection. 

It should be noted that state (C)  does not actually arise in the first, most unstable eigenvalue crossing of the $k_z=2$ linear instability but rather through a secondary appearance of a new unstable direction of $u_{lam}.$ Equivalently, the associated values of $k_x$ do not correspond to the  crossing of the main neutral curves plotted in figure \ref{fig:stab}. We therefore conjecture that further bifurcations at the first appearance (as $\alpha$ increases) will either give rise to yet more coherent nonlinear states, or reconnect to the states discussed here. 
However, our objective is not to conduct 
an exhaustive bifurcation analysis, 
but rather to identify the  origins of the converged chevron-shaped states which are of interest due to their
clear similarity to the previously reported zig-zag structures. Practically, we have also demonstrated the utility of continuation in $\alpha$ as it conveniently establishes the bifurcation structure of the various solution states. 

By definition all of the states in figures \ref{fig:BIF} and \ref{fig:BIF2}, with the exception of state (c), plotted by the green curves, result from the influence of stratification on the flow. The vertical component of the flow for these states is relatively weak compared to the horizontal component and as such the displacement of isopycnals is small, even though vertical gradients are significant. We omit figures for the sake of brevity but make the observation that the density layering observed in figures \ref{fig:DNS1}-\ref{fig:lu} requires additional shear instabilities to mix the density field; the flow field snapshots show clear Kelvin-Helmholtz-like overturning events (also see movie in the supplementary material). 

In figure \ref{fig:BIFB},
we show how the properties (in particular the streamwise velocity structures) 
of the state (A) change under
continuation in the stratification parameter $B$ for $\alpha$ fixed at its original value of $0.9$ and 
$Re=15$. After some turning points, the curve traced by the solution state  closes on itself.
Although it is  at least plausible that  continuation in  $B$ might lead to some smooth variation of the shear inclination as $B$ increases, it is actually apparent  that the increasing vertical structure observed in the numerical simulations arises due to the generation through instability of a  family of new invariant solution states. It is somewhat surprising that given the complicated behaviour of the solution state curve there is such little variation in the qualitative shape of the flow state, as visualised in figure \ref{fig:BIFB} by the streamwise velocity structure in a $(y,z)$ midplane at $x=\pi$. 
Indeed,  the direct numerical simulations show that these coherent structure states are subject to additional pattern forming instabilities, generically leading to  localisation of the shear in the $z$-direction and therefore turbulent motions.  Typical dynamical behaviour  is shown in figure \ref{fig:D5_3D} where three-dimensional visualisations of the streamwise velocity  $u$ are shown for the extremely strongly stratified and throttled simulation D5 (with $B=2000$) from table \ref{tab:DNS}. How such localisation comes about is a question for future research, but we note in passing, that although the notional value of $Re_B$ for this simulation is quite low {($Re_B=4.7$), it is still above 1} and the shear instability and layered flow structures bear more than a passing qualitative similarity to those shown in the vigorously turbulent and anisotropic 
simulation D9.6 of \cite{Brethouwer2007} (see in particular their figure 5a).

\begin{figure}
\begin{centering}
\includegraphics[width=\textwidth]{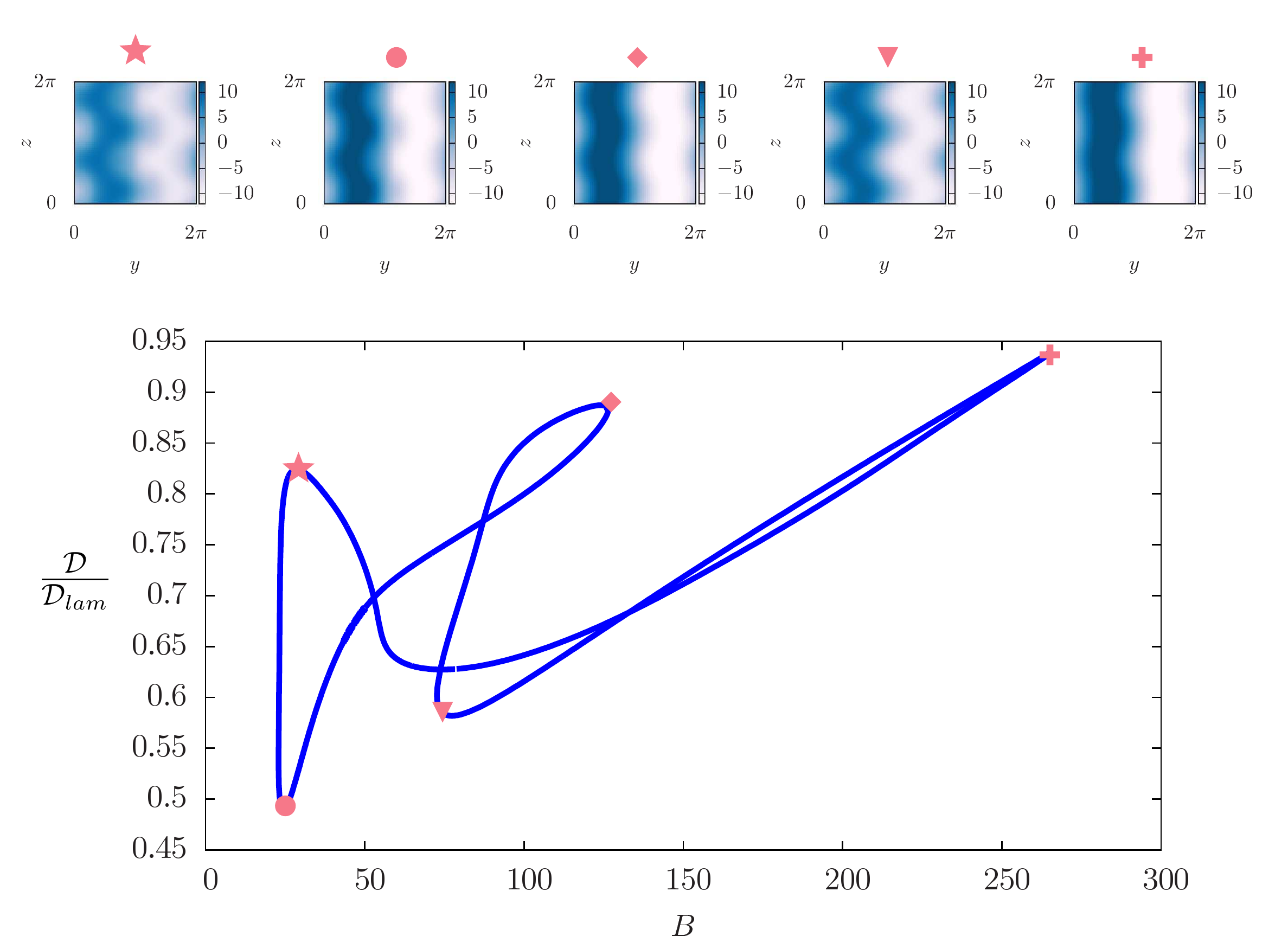}
\end{centering}
\caption{\label{fig:BIFB} Bifurcation diagram showing solution state branch (A)  from figure \ref{fig:BIF2} continued   in $B$ against normalised total dissipation $D/D_{lam}$ for flows at fixed $\alpha=0.9$ and $Re=15$. Visualisations of the streamwise velocity in the $y-z$ mid-plane at ($x=\pi$) at various labelled locations on the curve are shown in the  upper panels. Note that at large $B$ the solution curve bends back upon itself to close the loop. }
\end{figure}

\begin{figure}
\begin{centering}
\includegraphics[width=\textwidth]{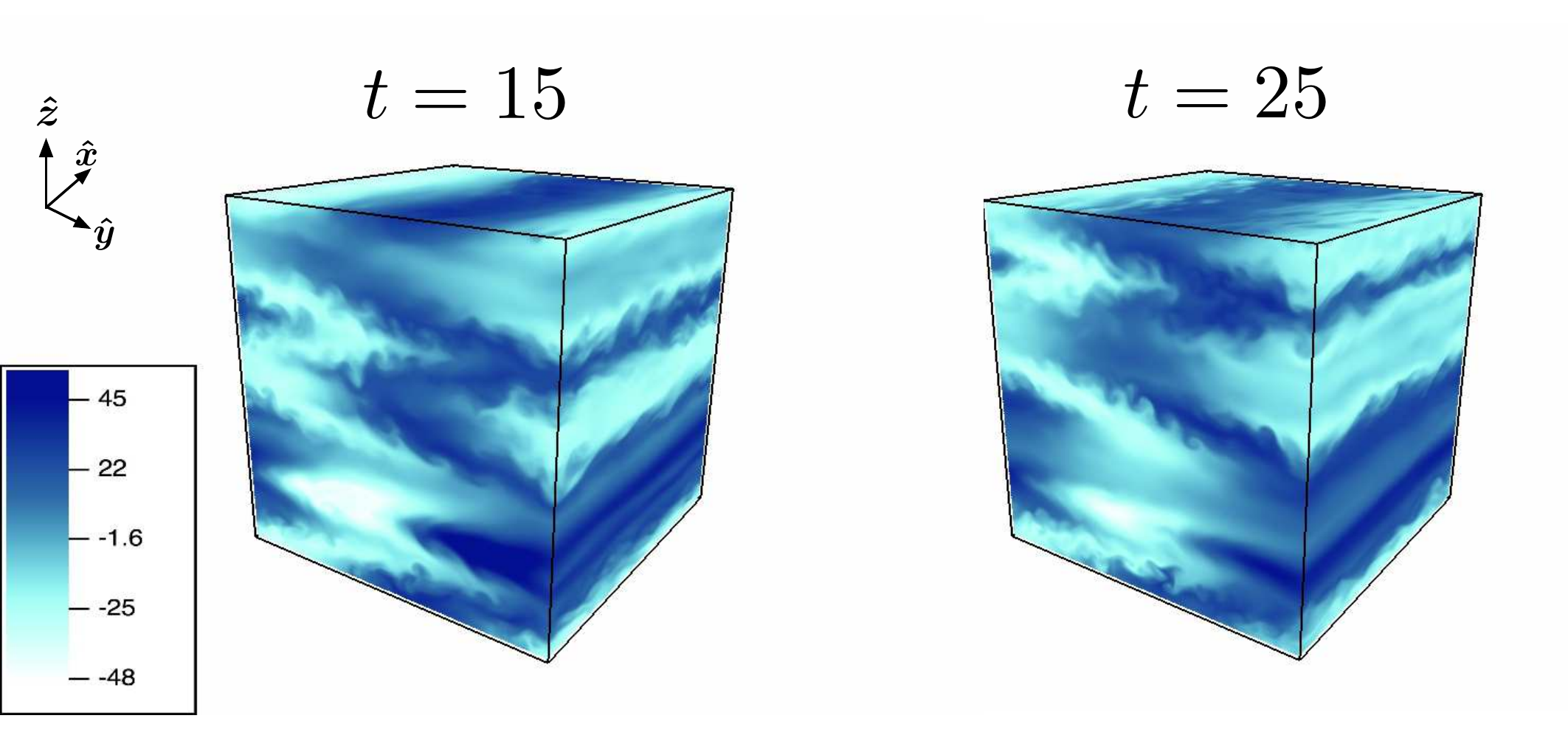}
\end{centering}
\caption{\label{fig:D5_3D} Snapshots of streamwise velocity, $u,$ for simulation D5 ($Re=50,\, D_0=250,\, B=2000$) showing the localisation in the vertical of the coherent structures leading to  inherently anisotropic and layered turbulent motions.  }
\end{figure}
%%%%%%%%%%%%%%%%%%%%%%%%%%%%%%%%%%%%%%%%%%%%%%%%%%%%%%%%%%%%%%%%%%%%%%%%
\section{Discussion}\label{sec:disc}

In this paper we have attempted to lay out the instability and bifurcation mechanisms by which the density field is spontaneously arranged into sustained layers by stratified turbulence driven with a horizontal shear. The key step is that we have been able to connect the coherent structures, observed to organise the mean flow of the turbulence into inclined shear layers, to the stratified linear instabilities of the basic horizontally varying and vertically uniform velocity profile. This is achieved by obtaining, directly from direct numerical simulations, various nonlinear coherent structures as  exact, yet unstable, steady states. Specifically, we have found  two steady states  which are striking representations of the mean flows observed during the turbulence, one state having a single vertical layer (\,(a1-3) in figure \ref{fig:BIF}) and the other state having two layers (\,(A) in figure \ref{fig:BIF2}). By constructing a bifurcation diagram using arc-length continuation in flow parameters, both are found to originate from a sequence of instabilities, the principal or primary  of which is a stratified linear instability of the base flow.

We have shown that this sequence of instabilities is responsible for breaking various symmetries of the flow. The bifurcations that we have isolated first break the $z$ and $x$ continuous invariance of the base flow through the linear instability studied in section \ref{sec:lin}. The ensuing primary states retain a discrete translational invariance $(x,y,z) \rightarrow (x+\pi/k_x,y,z+\pi/k_l)$ where $k_l$ is the vertical wavenumber of the instability. It is the breaking of this symmetry which leads to the formation of the chevron pattern with the same vertical wavelength as the primary state i.e.  $2\pi/k_l$ observed here. This is the salient distinction between primary states (b) or (C) and the states (a1)-(a3) or (A).  Importantly, our results do not exclude the possibility that other points of linear instability of the basic velocity profile could lead directly to a nonlinear saturated chevron state of the same vertical wavelength if $k_x=0$ at the bifurcation point or with half the vertical wavelength if $k_x \neq 0$ at the bifurcation point.  However, we have just not observed such primary bifurcations to chevron states here.

It should be emphasised that while the DNS calculations presented here fall at the periphery of the asymptotic scaling regimes {and also do not exhibit the anisotropy between horizontal and vertical length scales} traditionally considered when discussing ``stratified turbulence,'' particularly the LAST regime, this flow still shows significant modification of its mean profiles when, e.g. $Re_B$ is small and $F_h$ is relatively large. For example, case A3, where $Re_B=6.8$ and $F_h=0.074,$ shows very distinctive inclined layers in its mean flow (figure \ref{fig:DNS1} and table \ref{tab:DNS}). We should therefore stress that even when turbulence is considered to be weakly affected by stratification, the nonlinear flow response to buoyancy may still be significant. Such effects are possibly often masked by complicated stochastic forcing mechanisms \citep{Maffioli2016,2004JFM...517..281W,2014PhRvE..89d3002R}.

We expect the scenario outlined above to be quite general. For example, we conjecture the forcing wavenumber $n$ is not likely to change our analysis greatly, other than to adjust the linear stability regimes slightly and obviously add additional horizontal structure. The linear instability reported here is clearly closely related to
the three-dimensional stratified instabilities identified by  \cite{Billant:2000wg} for dipoles,  and \cite{DELONCLE:2007jl} for hyperbolic tangent shear layers and Bickley jets. Importantly, we have, for the first time, identified  a family of nonlinear \emph{steady} states which do indeed manifest a chevron-shaped or zig-zag vertical structure, very reminiscent of previously observed finite amplitude structures for a vortex dipole \citep{Billant:2000df,Billant:2000jg}. {Also in contrast to \cite{Billant:2000wg} and \cite{DELONCLE:2007jl} we have carried out a viscous analysis, and find that the scalings apparently set by an inviscid instability mechanism are still observable at finite amplitude and in situations where viscosity is playing a significant role.It is clearly of interest to investigate how robust this observation is, as the effects of nontrivial viscosity on stratified turbulence is often complex and subtle (see for example the discussion in \cite{Khani:2016}) }. Turbulence forced with these other horizontal profiles, therefore, may be expected to behave similarly, being organised about unstable steady states with vertically inclined shear layers. How such structures behave as the imposed horizontal shear becomes progressively more localised in space is an interesting and relevant research problem. 

The effect of inclination of initial shear relative to the background stratification has been considered in the freely decaying case by \cite{Jacobitz:1998di}. They found that even a small inclination angle is sufficient to increase turbulent production significantly. The effect of inclination angle on the linear instability of a Bickley jet was considered by \cite{Candelier:2011fb}, who found that growth rates are proportional to the sine of the angle. We therefore conjecture that a body-forced inclined shear will generate projected versions of the coherent structures discovered here, eventually vanishing as the shear approaches a purely vertical orientation. 

A further connection should be made to flows which are not driven by a large scale mean shear, but are known to exhibit layer formation, {including the $U/N$ scaling for layer depth}. Examples include oscillating grid or rod turbulence \citep{Holford:1999uc, 1999DyAtO..30..173H,Browand:1987cs} or the extensive literature on wakes, see for example \cite{Spedding:2002cl,DIAMESSIS:2011jy,Spedding:2014ez}. In such cases it is not uncommon to observe turbulent flows developing layers at relatively large Froude numbers ($\mathcal{O}(1),$ though of course caution must be exercised in the comparison of specific numerical values of parameters defined in different ways in different studies). Furthermore, at the moment, the precise mechanisms at play are still unclear. Several authors have discussed the possibility of a zig-zag mechanism in these flows \citep{Spedding:2002cl,Thorpe:2016ga}, which, given our findings that nonlinear saturated states underlying the layer pattern formation have their ultimate origins in such instabilities, increases the possibilities that other, more complicated base flows, than the one considered here, may also be subject to such behaviours. {Note that in the case of wakes some energy injection would be required, for example forcing with a vertically and horizontally varying streamwise flow, to locate steady nonlinear ECSs analogous to those presented here and connect them to a linear stability mechanism.}

One open question is how precisely the nonlinear solutions obtained correspond to episodes in the direct numerical simulations  at other parameter values, in particular at higher values of the Reynolds number. We do not address this question here, as our principal  aim is to demonstrate that  flow \emph{structures} of interest can not only be  located but also  their source (through an identifiable sequence of instabilities and bifurcations) can be determined. It is reasonable to expect that, given sufficient numerical resources,  corresponding states at a desired point in parameter space could also be found and we conjecture that a similar picture will emerge there. However it is also plausible that the coherent structures observed in one particular part of parameter space are mere `ghosts' of solutions existing in a nearby region of parameter space, and therefore solution efforts would fail to converge at that particular part of parameter space.  For example, $\alpha=1$ appears to be inaccessible to  most of the solution state curves described here, yet the coherent structures are still observed.

Significantly, the secondary and tertiary instabilities of the exact states we have identified support turbulence which rearranges the density field into well-defined mixed regions and interfaces which are sustained for long times. This aspect, and the mixing properties of this flow will be the subject of a subsequent paper. 
A fundamental question in stratified turbulence has been how, or indeed if, a stably stratified buoyancy field becomes organised into layers and interfaces with relatively deep mixed regions separated by relatively thin regions with strong gradients, and in particular how the vertical length scales of such layers might be set. Here we have shown, for the first time, a robust connection which extends all the way  from the linear instability of a simple laminar state, to the nonlinear saturated layered state. These nonlinear states which have been isolated are in striking qualitative agreement with the mean flows which are observed in the `turbulent flow' (in the sense that the flow is chaotic and exhibits a range of scales) at larger $Re$. As such it places the proposed scaling for the layer depth $l_v\sim U/N$ (in terms of a characteristic velocity scale $U$ and characteristic buoyancy frequency $N$) on a much stronger foundation.  The conventional but formally unjustified assumption that linear theory can explain some aspect of a highly nonlinear turbulent flow is actually here given some justification. However, it is still unclear how the basin of attraction for the turbulence is continuously deformed toward the observed layered configurations in phase space as the stratification is increased and how this relates to the appearance of the new unstable invariant manifolds (generated via the instabilities) about which the turbulence organises. 

\vspace{2cm}
\noindent
{\em Acknowledgements}. 
We extend our thanks, for many helpful and enlightening discussions, to Paul Linden, John Taylor, Stuart Dalziel and the rest of the `MUST' team in Cambridge and Bristol. We also thank the three anonymous referees whose constructive comments have significantly improved the clarity of the manuscript. The source code used in this work is provided at \url{https://bitbucket.org/dan_lucas/psgpu} and the associated data including initialisation files and converged states is found at \url{https://doi.org/10.17863/CAM.13090} This work is supported by EPSRC Programme Grant EP/K034529/1 entitled `Mathematical Underpinnings of Stratified Turbulence'. The majority of the research presented here was conducted when DL was a postdoctoral researcher in DAMTP as part of the MUST programme grant.
%%%%%%%%%%%%%%%%%%%%%%%%%%%%%%%%%%%%%%%%%%%%%%%%%%%%%%%%%%%%%%%%%%%%%%%%
\appendix
\section{DNS}
%%%%%%%%%%%%%%%%%%%%%%%%%%%%%%%%%%%%%%%%%%%%%%%%%%%%%%%%%%%%%%%%%%%%%%%%

\subsection{Recurrent bursts}
\begin{figure}
\begin{center}
\scalebox{0.65}{\input{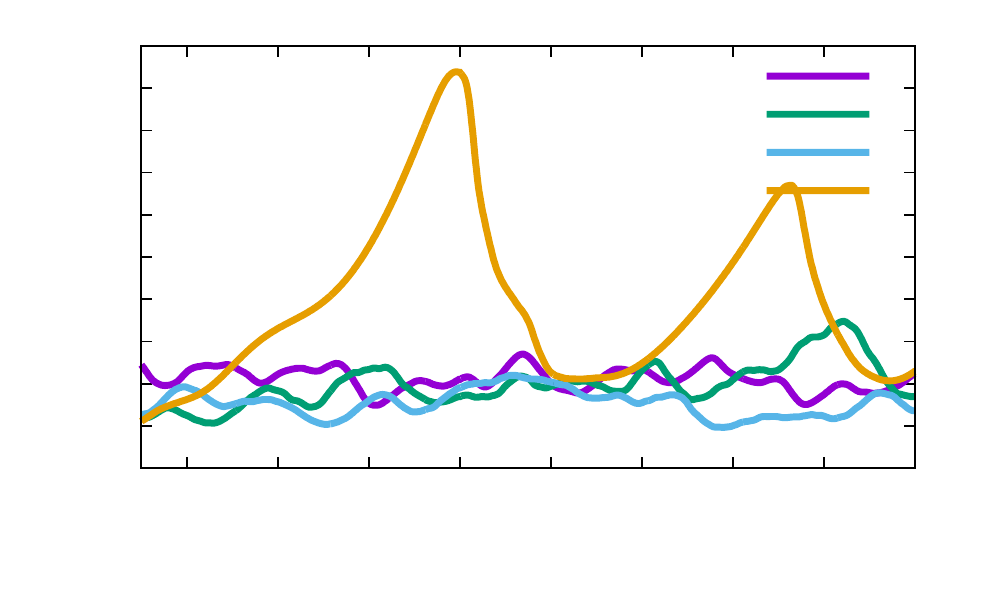}}
\scalebox{0.65}{\input{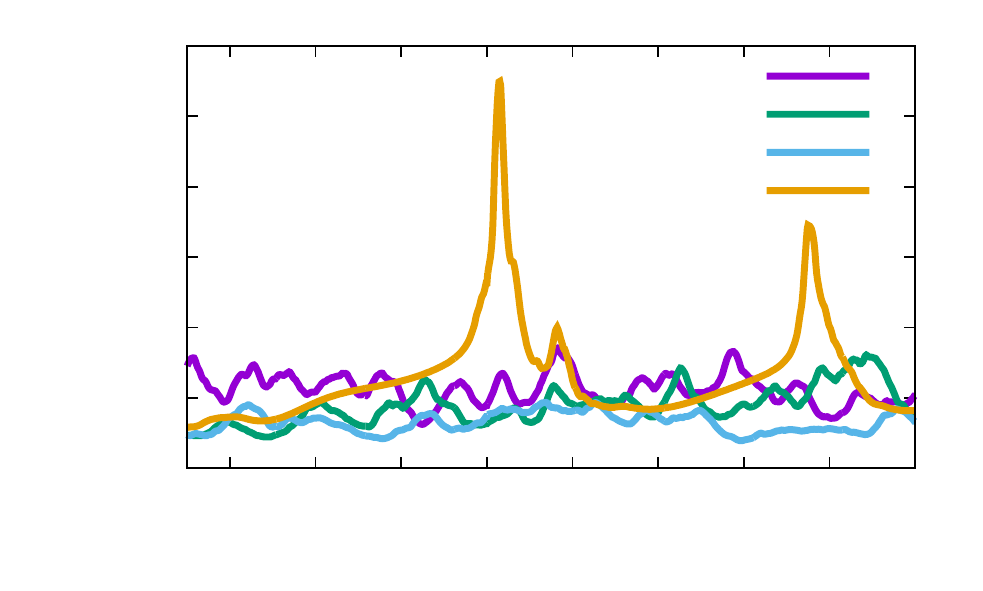}}
\end{center}
\caption{Time series of total kinetic energy density $\mathcal{K}=(1/2)\langle \bm |u|^2\rangle_V$ (left panel) and volume-averaged scaled dissipation rate $\mathcal{D}/\mathcal{D}_{lam}=2\langle |\nabla \bm{u}| ^2\rangle_V/Re^2$ (right panel),  
for simulations A1-A4 from table \ref{tab:DNS}. Note the widely separated and yet intense recurrent  bursts for  simulation A4 with $B=50$, and the lag
between $\mathcal{K}$ and the ensuing dissipation $\mathcal{D}$. \label{fig:DE1}}
\end{figure}

\begin{figure}
\begin{center}
\includegraphics[width=0.9\textwidth]{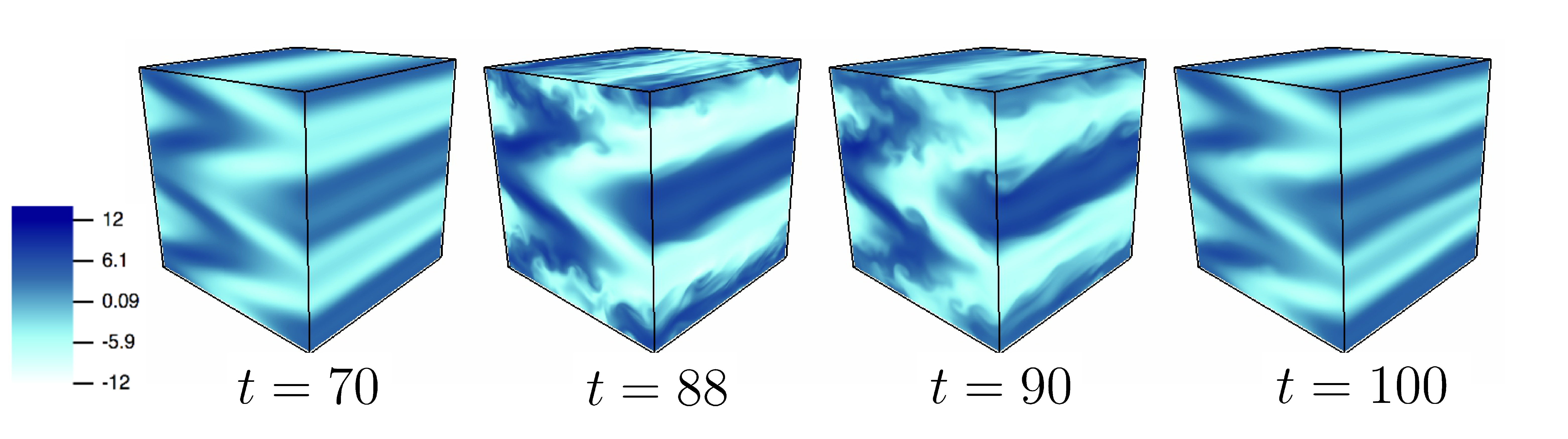} 
\end{center}
\caption{Three-dimensional snapshots of streamwise velocity $u$ at times $t=70,\,88,\,90,\,100$ showing the burst of turbulence for simulation A4 as defined in table \ref{tab:DNS} with $Re=100$, $B=50$, $Pr=1$, $\alpha=1$. \label{fig:Ri50_3D}}
\end{figure}

One important feature of the inclined shear, observed from the DNS and figure \ref{fig:DNS1}, is that the mean flow now has a component of vertical shear. We find that for simulations with {fixed $Re$ and} large $B$ (i.e. for $ B\geq 50$) vertical shear instability (of Kelvin-Helmholtz (KH) type) is suppressed and density variations with respect to $z$ remain small. Due to the body-forced nature of this flow, quiescent episodes, when the vertical shear instability is suppressed, are associated with a build up of energy in the mean flow as energy is continually and uniformly injected via the forcing. This results in highly intermittent turbulent bursts. The flow is accelerated until apparently a local gradient Richardson number criterion is overcome, KH shear instability is initiated which then grows to finite amplitude, breaks down to disordered motion and then decays until the mean is regenerated once again by the forcing. In fact this is the generic behaviour observed when the forcing itself provides vertical shear, i.e. $\sin(nz)\hx$ (figure not shown).  The final row of figure \ref{fig:DNS1} shows some turbulent remnant in the $T=50$ snapshots with apparent shear instability on the vertical bands of $u.$ Figure \ref{fig:Ri50_3D} shows the full three-dimensional streamwise velocity for four snapshots in time, spanning the burst of spatiotemporal turbulence around $t \simeq 80-90$ apparent in figure \ref{fig:DE1}.

\begin{figure}
\begin{center}
\scalebox{0.55}{\input{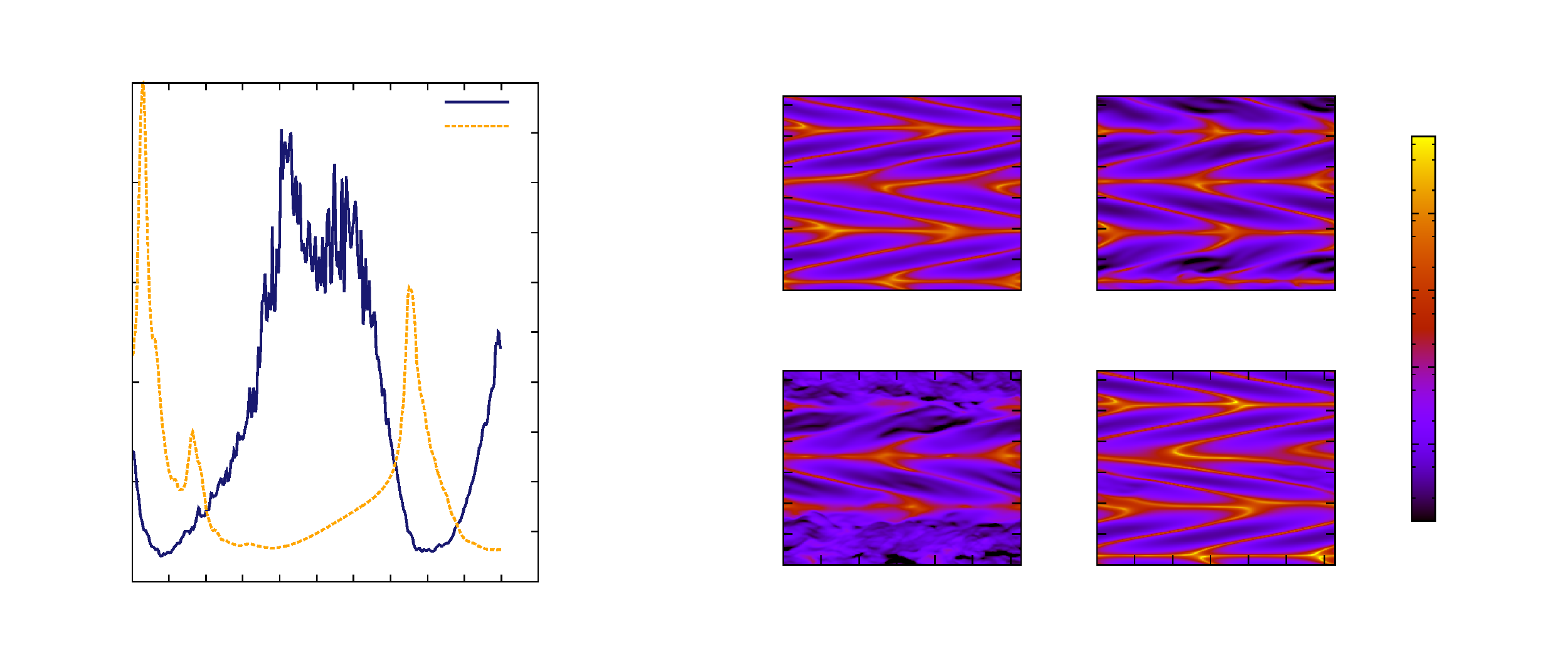}}
\end{center}
\caption{Left: Time dependence of the volume average of $Ri_G$ 
and the scaled dissipation rate $\mathcal{D}/\mathcal{D}_{lam}$, 
showing the anti-correlation of turbulence intensity and gradient Richardson number;
Right: Four snapshots of $Ri_G(y,z,t)$ at times $t=70,\,88,\,90,\,100$
from simulation  A1 with $Re=100$, $B=50$, $Pr=1$, $\alpha=1$. 
Note the small values of $Ri_G(y,z,t)$ (plotted as dark colours) in the panels for $t=88$ and $t=90$ associated with the burst of turbulence
near the top and bottom of the domain. By $t=100$ the stabilising lattice structure of high values
apparent in the panel for $t=70$ has become re-established. Angled strips of high $Ri_G$ are
associated with minimal vertical shear, and horizontal strips of high $Ri_G$ are associated with maximal
density gradient.
\label{fig:RiG}}
\end{figure}

Defining an appropriate gradient Richardson number
\begin{equation}
 Ri_G(y,z,t) = \frac{-B\left\langle\frac{\partial\rho_{tot}}{\partial z}\right\rangle_x}{\left\langle\left(\frac{\partial u_h}{\partial z}\right)^2\right\rangle_x},
\label{eq:rigdef} 
\end{equation}
where the subscript denotes averaging in the streamwise $x-$direction, 
we plot a time series of the volume average $\langle Ri_G\rangle_{V}$ and $y-z$ snapshots at various times  for the $B=50$ simulation in figure \ref{fig:RiG}.  We find that the suppression of shear instability occurs in two distinct classes of thin strips. One class (at a non-trivial angle to the vertical) is  aligned with the minimum of the vertical shear. The other class (close to horizontal) is associated with the maximum of the density gradient, as is  apparent from the bottom right two panels in figure \ref{fig:DNS1}.  In other words, the horizontal structures in figure \ref{fig:RiG} are caused by  large values of density gradient in the numerator of $Ri_G$ and the inclined structures are caused by minima of the shear in the denominator of $Ri_G$. This lattice structure of the two classes of strips effectively covers the entire flow domain, and eventually manages to  stabilise the flow globally. Once the flow is accelerated again sufficiently, we then observe instability nucleating in the regions of minimal $Ri_G$  in the gaps of this lattice structure, as is apparent in the  frames associated with $t=88$ in figures \ref{fig:Ri50_3D} and \ref{fig:RiG}. Overall, for this flow,  the volume average of $Ri_G$ reaches a maximum value of  $Ri_G\approx 0.2$ during the stabilising period, but it is not clear whether there is any significance to this particular value.

From a dynamical systems point of view, one may consider these observed cycles to be similar to the type of homoclinic tangle studied to explain bursting in plane Couette flow by \cite{vanVeen:2011cm}. We appear to observe a close approach to a coherent structure and a return to it via a complex turbulent trajectory which wraps up the stable and unstable manifolds of the underlying solution. Unfortunately, this particular feature of close approach and complex (near) return makes this flow quite challenging computationally. As is particularly apparent for simulation A4 as shown in figure \ref{fig:DE1},  as the buoyancy parameter $B$ is increased further (i.e. the background buoyancy frequency is increased relative to the horizontal shear forcing)  the mean flow required to maintain turbulence becomes stronger with more intense small-scale dissipation. This `strengthening' of the turbulence requires both smaller timesteps to maintain numerical stability as the mean flow increases, and also longer integrations to approach steady statistics and average out the increasingly long intermittent bursting time scale. 

%%%%%%%%%%%%%%%%%%%%%%%%%%%%%%%%%%%%%%%%%%%%%%%%%%%%%%%%%%%%%%%%%%%%%%%%

\subsection{Throttling}
To attempt to overcome the combined computational challenge of increasingly long times between increasingly intense intermittent bursts as $B$ is increased, we employ a throttling method similar to that used by \cite{2012JFM...696..434C}. This throttling modulates the forcing amplitude with the aim of maintaining a mean dissipation rate in the vicinity of a target value. 
Therefore, we allow the forcing term in equation (\ref{NSu}) to have a time varying amplitude, i.e.  
\beq \bm f = \chi(t)\sin(ny)\hx. \label{eq:chit}
\eeq
We then adjust $\chi(t)$ based on the instantaneous kinetic energy budget, i.e.
\beq
\frac{\ud \mathcal{K}}{\ud t} = \mathcal{I} + \mathcal{B}-\mathcal{D}\label{eq:KE}
\eeq
where $\mathcal{K},$ $\mathcal{I}$, $\mathcal{B}$ and $\mathcal{D}$ are the total kinetic energy density, volume-averaged energy input, buoyancy flux and dissipation rate respectively, as defined in (\ref{eq:diag1})-(\ref{eq:diag2}).
Following \cite{2012JFM...696..434C} we set a target dissipation rate $\mathcal{D}_0$ and seek $\frac{\ud \mathcal{K}}{\ud t}=0,$ i.e. statistically stationary kinetic energy, so that
\beq 
 \chi(t) = \frac{\mathcal{D}_0-\mathcal{B}}{\langle u \sin(y) \rangle}.
\label{eq:chit2}
\eeq
In other words when $\mathcal{D}<\mathcal{D}_0$ the energy input is increased, so that the total kinetic energy grows in time, and when $\mathcal{D}>\mathcal{D}_0$ the flow is decelerated by reducing $\mathcal{K}$. Notice that the connections between the terms in equation (\ref{eq:KE}) are  inherently nonlinear and linked to the turbulent cascade. Energy is input at the largest scale and is ultimately dissipated at much smaller scales. For this reason there is an inevitable lag between the adjustment of $\chi$ and the flow response. Furthermore, there are still undoubted computational challenges.
To maintain adequate resolution at large external $Re$ we cannot throttle with large $\mathcal{D}_0$, although for statistical stationarity, larger values of $B$ require progressively larger $\mathcal{D}_0$. For these reasons we choose the parameters in table \ref{tab:DNS}  to maintain at least some semblance of a  turbulent state at progressively stronger stratification.

\begin{figure}
\begin{center}
\scalebox{0.45}{\input{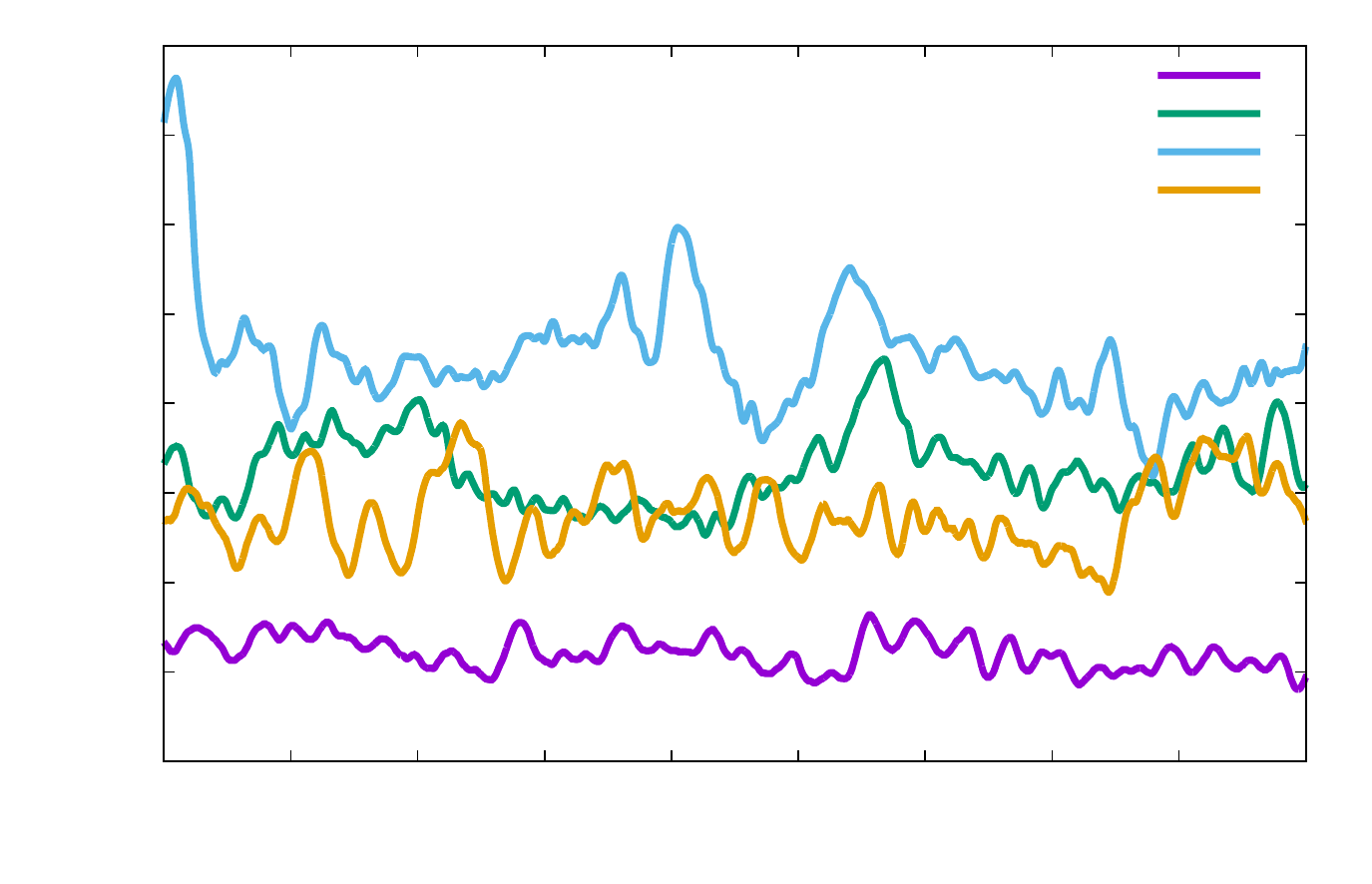}}
\scalebox{0.45}{\input{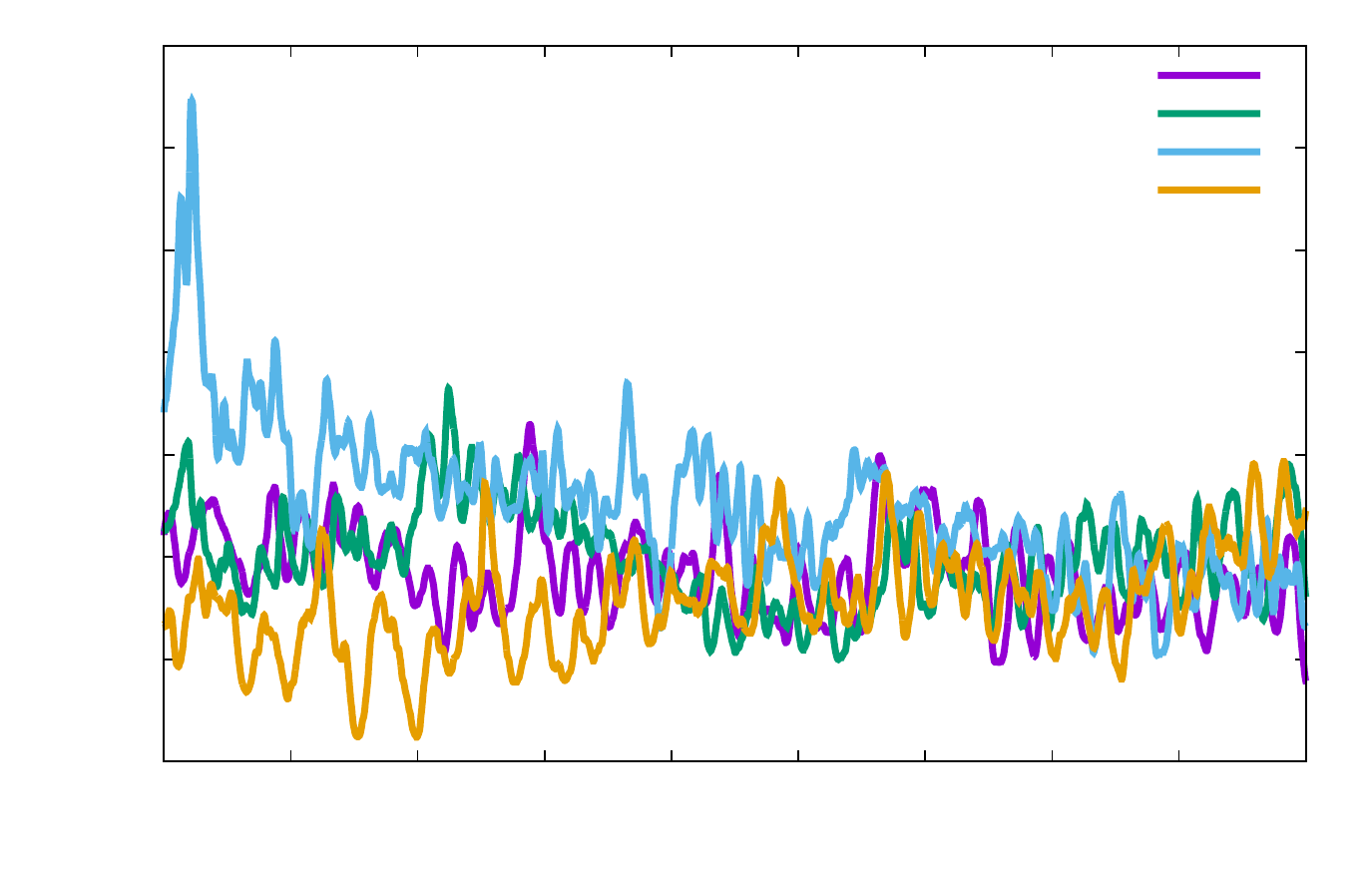}}
\end{center}
\caption{Time dependence of total kinetic energy $\mathcal{K}=(1/2)\langle |\bm u|^2\rangle_V$ 
and scaled dissipation rate $\mathcal{D}/\mathcal{D}_0=\langle |\nabla \bm{u}| ^2\rangle_V/Re^2$ for the various throttled simulations at large $B$ whose fields are plotted in figure \ref{fig:throttle1}. Note that the  dissipation fluctuates near, yet systematically below the target value $\mathcal{D}_0$. \label{fig:DEthrottle}}
\end{figure}

Figure \ref{fig:throttle1} shows equivalent  $yz-$plane $\rho$ and $u$ snapshots and means to those shown in figure \ref{fig:DNS1}  for the throttled simulations D1,  B3, C2 and  C3 as listed in table \ref{tab:DNS}, 
associated with larger values of the buoyancy parameter $B$.  Figure \ref{fig:DEthrottle} shows the 
time dependence of the kinetic 
energy density  $\mathcal{K}$ and scaled dissipation rate $\mathcal{D}/\mathcal{D}_0$ for these throttled simulations.
The throttled simulations  exhibit approximately statistical stationarity near the target dissipation rate $\mathcal{D}_0$ in contrast to the observed time-lagged bursting in both energy density and dissipation rate shown   in figure \ref{fig:DE1} for the unthrottled simulation A4 with $B=50$. 
Curiously, the throttling method doesn't reach the target dissipation rate $\mathcal{D}_0$ as closely as reported in \cite{2012JFM...696..434C}, with 
a systematic undershoot in the actually occuring dissipation rate $\mathcal{D}$. This 
difference is presumably due to the different nature of forcing between the two studies.  In particular,  \cite{2012JFM...696..434C} enforce a specific background mean linear shear rather than a `free' body force as in our simulations. In spite of this discrepancy, 
the method serves our purpose by providing a much more stationary flow at large $B$ and the opportunity to investigate the trend in layer scale with $B$.

\begin{table} \setlength{\tabcolsep}{8pt}
\begin{center}
\begin{tabular}{cccccccccccc}
\#&$Re$& $B$ & $D_0$  &$U$ &  $\eta k_{\mathrm{max}}$& $ l_{\bar{u}}$ & $l_o$ &$F_h$ & $Re_B$  & $Re_\lambda$ \\
\hline
A1&100&1	&-&  	 1.3&	 2.6&		2.5 & 1.05 & 0.66 &110.9 &56.6 \\
A2&100&5	&-&	1.9&	2.7&		3.9 & 0.3 & 0.13 &20.2  & 48.6 \\
A3&100&10	&-&	1.7&	3.0&		3.5 & 0.15 &0.074 &6.8 & 41.8 \\
A4& 100&50	&-&	3.1&	2.5&		1.8 & 0.06 & 0.021&2.9  & 36.8\\
\hline
B1& 100 & 50 & 50  & 7.3&1.03 & 4.5 & 0.37 & 0.126 &94.9 & 178.7 \\
B2&100 & 100 & 50 & 5.9  &1.1 & 2.8 & 0.18 & 0.099 & 34.3 & 204.0 \\
B3&100 & 200 & 50 & 6.9 & 1.08 & 2.1& 0.12 & 0.057 &19.3 & 159.8 \\[4pt]

C1&75 & 500  & 100  & 6.3 &1.2 & 1.6& 0.073 & 0.068 &9.0 & 263.8\\
C2&75 &1000 & 100 & 9.8& 1.1& 1.2&0.052 &0.028&6.4& 107.0 \\
C3&75 &2000 & 100  & 11.5 &1.1&1.2 &0.03&0.014&3.02&  55.6 \\[4pt]

D1&50 & 500  & 100  & 7.8&1.6 & 1.7 & 0.079 &0.052 & 7.1& 54.3 \\
D2&50 & 200 & 250  & 11.0 & 1.2 & 4.4 & 0.26 &0.115&49.4&  89.2\\
D3&50 & 500 &  250 & 10.8&1.3 & 2.8&0.11 & 0.057&14.9 & 114.2\\
D4&50 &1000 & 250 & 10.7&1.3 & 2.4& 0.071 &0.044& 7.9 & 121.2 \\
D5& 50 &2000 & 250 & 12.1 &1.2 & 1.4&0.046 &0.029&4.7 & 131.2\\

\end{tabular}
\end{center}
\caption{\label{tab:DNS} Imposed parameters and diagnostic simulation outputs for the four groups of simulations.  Simulations in group A are unthrottled simulations, simulations in group B are throttled simulations with $Re=100$, simulations in group C are throttled simulations with $Re=75$, and
simulations in group D are  throttled simulations with $Re=50$. The 
Kolmogorov microscale, 
within the chosen nondimensionalization is given by $\eta=\left(Re^3 \bar{\mathcal{D}}\right)^{-1/4}$
and is scaled by the  maximum wavenumber allowable in the  simulation $k_{\mathrm{max}}=N_x/3=85$. The nondimensional Ozmidov length scale $l_O=\left({\bar{\mathcal{D}}}/{B^{\frac{3}{2}}}\right)^{1/2}$, horizontal Froude number $F_h={D}/{\sqrt{B}U^2},$ the  buoyancy Reynolds number $Re_B={\bar{\mathcal{D}} Re}/{B},$ a typical horizontal velocity scale given by $U=\langle |\bar{u}| \rangle_{V}$ and the Taylor microscale Reynolds number $Re_\lambda=\mathcal{K}_{\mathrm{turb}}\sqrt{{10Re}/{\bar{\mathcal{D}}}}$ 
are also listed. Here we define $\mathcal{K}_{turb} = (1/2)\langle(\bm u - \bar{\bm u})^2\rangle_V$ and overbars are always time averages over the full $T$ window. Group A have $T=100$ and groups B-D $T=50.$}
\end{table}
%%%%%%%%%%%%%%%%%%%%%%%%%%%%%%%%%%%%%%%%%%%%%%%%%%%%%%%%%%%%%%%%%%%%%%%%
%\newpage

\bibliography{papers_dl,papers_cpc}
\bibliographystyle{jfm}

\end{document}